\documentclass[%
 english,
 reprint,
superscriptaddress,
showpacs,
 amsmath,amssymb,
 aps,
prb,
]{revtex4-2}

\usepackage[T1]{fontenc}
\usepackage[latin1]{inputenc}
\usepackage{graphicx}
\usepackage{color}
\usepackage{hyperref}
\hypersetup{
	colorlinks = true,
	allcolors = {blue}
}

\usepackage{natbib}

\usepackage{babel}
\makeatother

\begin{document}

\title{Rotational symmetry breaking at the incommensurate charge-density-wave transition in Ba(Ni,Co)$_2$(As,P)$_2$: Possible nematic phase induced by charge/orbital fluctuations}

\author{Michael Merz}
\email[Corresponding author: ]{michael.merz@kit.edu}
\affiliation{Institute for Quantum Materials and Technologies, Karlsruhe Institute of Technology, 76021 Karlsruhe, Germany}

\author{Liran Wang}
\thanks{Present address:\@ Kirchhoff Institute of Physics, Heidelberg University, INF 227, 69120 Heidelberg, Germany}
\affiliation{Institute for Quantum Materials and Technologies, Karlsruhe Institute of Technology, 76021 Karlsruhe, Germany}

\author{Thomas Wolf}
\affiliation{Institute for Quantum Materials and Technologies, Karlsruhe Institute of Technology, 76021 Karlsruhe, Germany}

\author{Peter Nagel}
\affiliation{Institute for Quantum Materials and Technologies, Karlsruhe Institute of Technology, 76021 Karlsruhe, Germany}

\author{Christoph Meingast}
\affiliation{Institute for Quantum Materials and Technologies, Karlsruhe Institute of Technology, 76021 Karlsruhe, Germany}

\author{Stefan Schuppler}
\affiliation{Institute for Quantum Materials and Technologies, Karlsruhe Institute of Technology, 76021 Karlsruhe, Germany}

\date{\today}

\begin{abstract}
Nematic phase transitions in high-temperature superconductors have a strong impact on the electronic properties of these systems. BaFe$_2$As$_2$,\@ with an established nematic transition around 137 K induced by magnetic fluctuations, and BaNi$_2$As$_2$,\@ a non-magnetic analog of BaFe$_2$As$_2$ with a structural transition in the same temperature range,\@ share a common tetragonal aristotype crystal structure with space-group type $I4/mmm$.\@ In contrast to BaFe$_2$As$_2$ where collinear stripe magnetic order is found for the orthorhombic low-$T$ phase, a unidirectional charge density wave together with (distorted) zig-zag chains are observed for the triclinic low-$T$ phase of BaNi$_2$As$_2$.\@ Here we show that between the high- and low-$T$ phases of Ba(Ni,Co)$_2$(As,P)$_2$ an additional phase with broken fourfold symmetry and $d_{xz}$ orbital order exists which is a promising candidate for charge/orbital-fluctuation-induced nematicity. Moreover, our data suggest that the enhanced $T_{\rm c}$ found for Ba(Ni,Co)$_2$(As,P)$_2$ for higher Co or P substitution levels might result from suppression of the (distorted) zig-zag chains by reducing the contribution of the $d_{xy}$ orbitals.\@
\end{abstract}

\maketitle

\section{Introduction}

High-temperature superconductivity in iron-based pnictides (FeSCs) \cite{Kamihara2008} has triggered strong interest in the layered FeSCs during the last decade.\@ The prototypical system BaFe$_2$As$_2$ (Fe122) is an antiferromagnet (AFM) with collinear stripe magnetic order below T$_{\rm N} \approx 134$ K.\@ Its first-order magnetic transition at T$_{\rm N}$ is preceded at slightly higher temperature, $T_{\rm nem}$,\@ by a second-order structural one from the tetragonal space group (SG) $I4/mmm$ to the orthorhombic SG $Fmmm$ (even more so if Fe is slightly replaced by Co or Ni) \cite{Kim2011a,Fernandes2014}.\@ The latter, so-called nematic, transition has an electronic origin and is (without a final smoking-gun proof \cite{footnote19}) often believed to be induced by magnetic fluctuations which break the fourfold symmetry and induce the nematic/structural transition at $T_{\rm nem}$ \cite{Fernandes2014} [see  Appendix \ref{AppD}].\@ While Fe122 is non-superconducting, superconductivity with a transition temperature $T_{\rm c}$ as high as $\approx 40$ K appears as a dome-like phase when T$_{\rm N}$ and $T_{\rm nem}$ are significantly suppressed --- either by (partial) substitution on the various sites (As, Ba, Fe) or by pressure \cite{Rotter2008b,Canfield2010,Colombier2009}.\@

BaNi$_2$As$_2$ (Ni122) is a non-magnetic superconducting analog of Fe122.\@ It shares the same tetragonal high-temperature (high-$T$) structure and exhibits a structural transition at a very similar temperature ($\approx 131$ K) \cite{Sefat2009,Kudo2012}.\@ However, no magnetic ordering is observed for the low-temperature (low-$T$) phase of Ni122.\@ Instead a commensurate unidirectional charge density wave (C-CDW) is found \cite{Sefat2009,Lee2019} and the structural transition is of first order from tetragonal (SG $I4/mmm$) to triclinic (SG $P$${\bar 1}$) \cite{Sefat2009,Kudo2012}.\@ 
A difference between the low-$T$ structures of the two compounds is that linear Fe-Fe chains are found for Fe122 and Ni-Ni zig-zag chains for Ni122 \cite{Sefat2009,Lee2019}.\@ In analogy to Fe122, where $T_{\rm c}$ can be increased by suppressing the long-range magnetic order, $T_{\rm c}$ can be enhanced for Ni122 from $0.6$ K to $\approx 3.6$ K upon substitution on the pnictide \cite{Kudo2012}, the Ni  \cite{Kudo2017,Lee2019}, or the Ba site \cite{Eckberg2019} by suppressing the C-CDW, i.e.,\@ the long-range charge order.\@ This raises fundamental questions regarding the analogies between spin and charge modulations, in particular with respect to electronic nematicity and their relevance for superconductivity.\@ And indeed, theoretical investigations point to charge fluctuations as a possible origin of the zig-zag chain formation and of the strong-coupling superconductivity observed in (partially) substituted Ni122 \cite{Yamakawa2013}.\@ Along the same lines, first experimental indications for a nematic fluctuation-enhanced $T_{\rm c}$ have, very recently, been reported for (Ba,Sr)Ni$_2$As$_2$ \cite{Eckberg2019}.\@

The most important prerequisites for charge/orbital-fluctuation-induced nematicity in Ni122 [see also Appendix \ref{AppD}] are as follows \cite{Fernandes2014}:\@ In the high-$T$ phase, the structure has tetragonal symmetry with SG $I4/mmm$.\@ Upon cooling, charge fluctuations increase and at $T_{\rm nem}$ they become strong enough to trigger a second-order nematic/structural transition. In a temperature range $T_{\rm CDW} \leq T \leq T_{\rm nem}$ of the nematic phase the system spontaneously develops an orbital order and the fourfold symmetry is broken \cite{Fernandes2014}.\@ Finally the translation symmetry is broken below $T_{\rm CDW}$ as well and a CDW ground state appears.\@

Here we will demonstrate that Ba(Ni,Co)$_2$(As,P)$_2$ is a promising candidate for such a charge/orbital-fluctuation-induced nematicity.\@ By combining the strengths of thermal expansion (TE), single-crystal x-ray diffraction (XRD),\@ and near-edge x-ray absorption fine structure (NEXAFS), we show that charge/orbital fluctuations exist far above the nematic transition, that the fourfold symmetry of the high-$T$ phase is broken below $T_{\rm S,1}$ (as will be shown our experimental $T_{\rm S,1}$ corresponds to $T_{\rm nem}$ in the theoretical model described above) most probably by these orbital fluctuations, and that an orbital order develops in the temperature range $T_{\rm S,2} \leq T \leq T_{\rm S,1}$.\@ This nematic phase is accompanied by an incommensurate charge density wave (I-CDW).\@ For the low-$T$ phase below $T_{\rm S,2}$ (as will be shown our experimental $T_{\rm S,2}$ corresponds to $T_{\rm CDW}$ in the theoretical model described above),\@ the ground state C-CDW, which develops directly from the I-CDW, appears together with the zig-zag chains.\@ Suppressing the formation of the (distorted) zig-zag chains and the accompanying C-CDW by reducing the contribution of the $d_{xy}$ orbitals seems to be responsible for the enhanced $T_{\rm c}$ observed for higher Co or As substitution levels.\@

\begin{figure}[t]
\hspace{-7mm}
\includegraphics[width=0.5\textwidth]{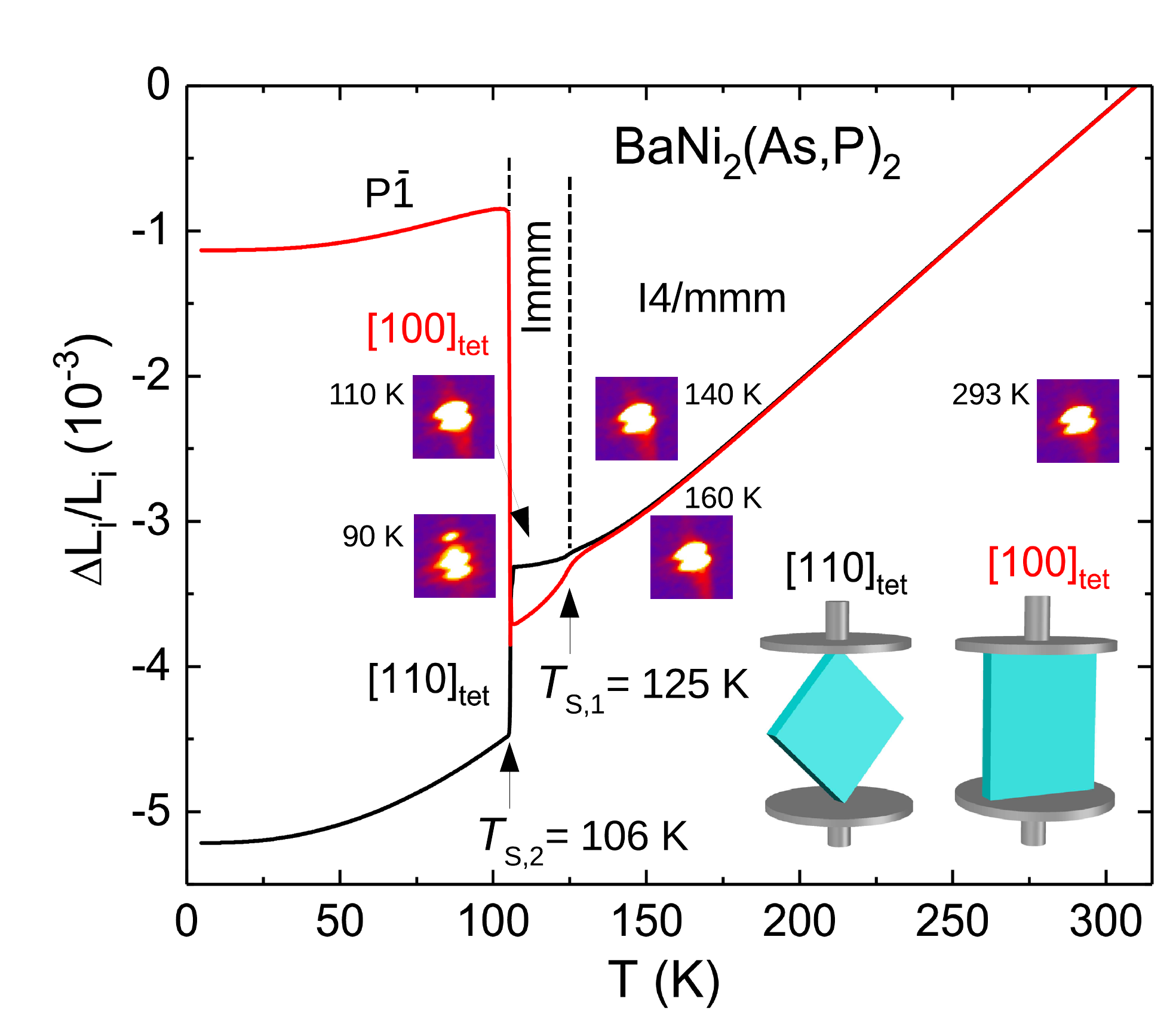}
\caption{\label{fig1} (Color online) Relative length change, $\Delta L/L$, versus temperature of the [100]$_{\rm tet}$ and the [110]$_{\rm tet}$ direction in the high-$T$ tetragonal notation for BaNi$_2$(As,P)$_2$ obtained using high-resolution capacitance dilatometry. The splitting of the (${\bar 3}$ ${\bar 3}$ ${\bar 6}$)$_{\rm tet}$ reflection at the structural phase transition to triclinic obtained from XRD measurements is shown as well. The observation of two peaks already for the tetragonal symmetry above $T_{\rm S,1}$ comes from the K$_{\alpha 1,2}$ doublet.}
\end{figure}

\section{Experimental\label{experimental}}

Single crystals for all compositions in this work were grown from self-flux and characterized using XRD and energy-dispersive x-ray spectroscopy as described in \cite{Merz2016}.\@ The resulting crystals typically have a surface with a size of about 3 $\times$ 2 mm$^2$ and a thickness around 0.5 mm.\@ The P content of the BaNi$_2$(As,P)$_2$ sample was determined to 3 \% and the Co content of the Ba(Ni,Co)$_2$As$_2$ crystal to 5 \%.\@ TE was measured using a high-resolution homebuilt capacitance dilatometer \cite{Meingast1990}.\@ XRD data were measured on pieces of the same samples used for TE.\@ Temperature-dependent x-ray diffraction data on Ni122, BaNi$_2$(As,P)$_2$,\@ Ba(Ni,Co)$_2$As$_2$\@ and Fe122 single crystals were collected between 90 K and room temperature on a STOE imaging plate diffraction system (IPDS-2T) equipped with Mo $K_{\alpha}$ radiation. NEXAFS measurements at the Ni $L_{2,3}$ edges were performed at the Institute for Quantum Materials and Technologies beamline WERA at the KARA synchrotron light source (Karlsruhe, Germany) on samples from the same batches in an analogous way as outlined in \cite{Merz2016,Merz2010,Merz2011,Merz2012}.\@

\section{Results and Discussion\label{results}}

\begin{figure}[b]
\hspace{-0.0mm}
\includegraphics[width=0.5\textwidth]{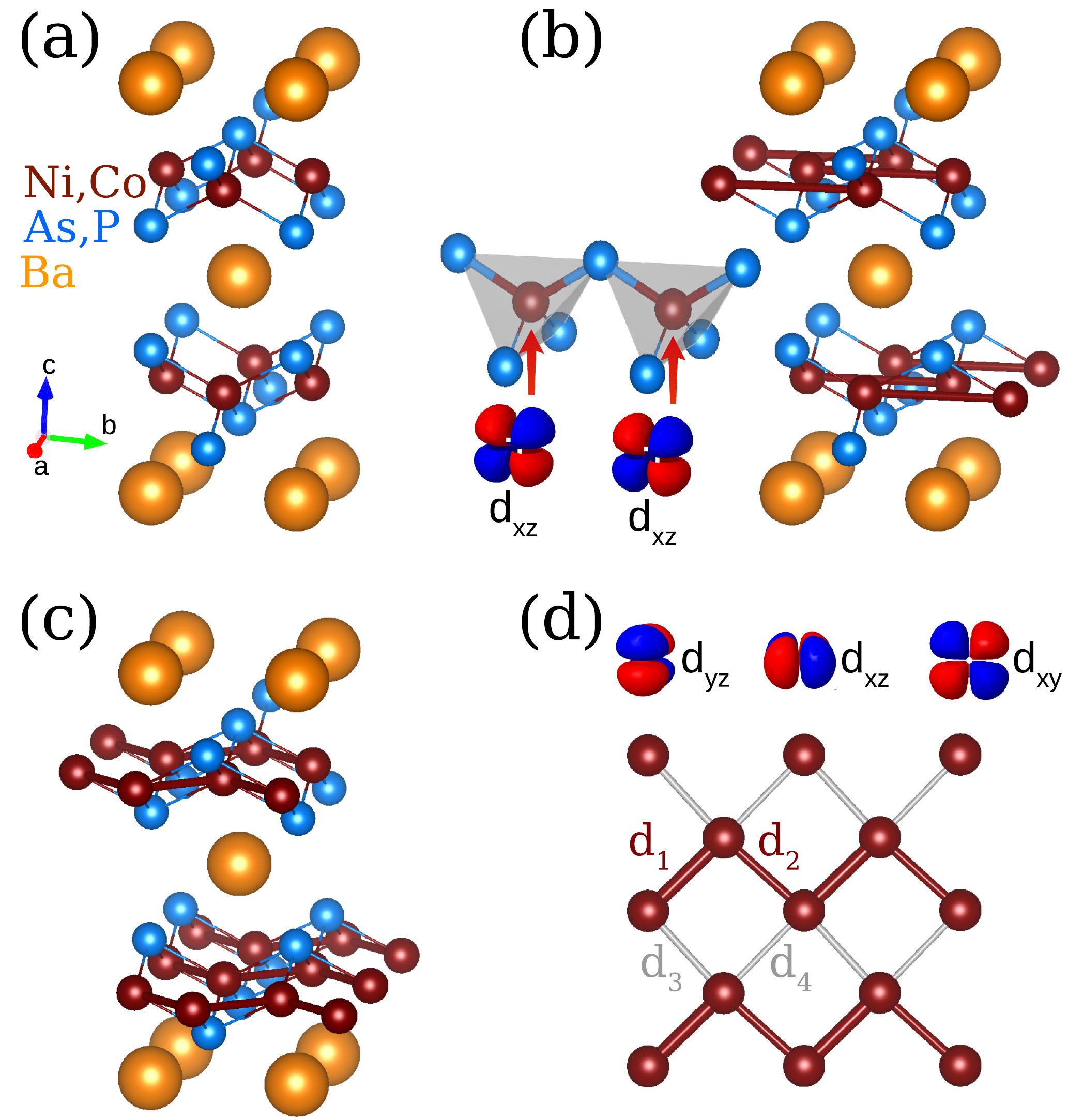}
\caption{\label{fig6}(Color online) (a) High-$T$ tetragonal spatial structure of BaNi$_2$(As,P)$_2$, Ni122, and Ba(Ni,Co)$_2$As$_2$. (b) Spatial structure of the orthorhombic/nematic phase in the range $T_{\rm S,2} \leq T \leq T_{\rm S,1}$.\@ The sketch of two adjacent tetrahedra illustrates how the reduced bond lengths (represented by thicker lines) lead to the staggered orbital ordering induced by the lifting of the degeneracy of $d_{xz}$ and $d_{yz}$ orbitals.\@ (c) Low-$T$ triclinic structure for which the Ni-Ni bonds form zig-zag chains or connected `Ni-Ni dimers' (see text). (d) Top view of a single Ni layer. The four different Ni-Ni bond distances which are induced by the symmetry reduction to triclinic are indicated in this figure by different line thicknesses together with the orientation of the relevant orbital types.}
\end{figure}
Fig.\@ \ref{fig1}(a) presents the high-resolution TE, $\Delta L/L$,\@ measured along the [100]$_{\rm tet}$ and the [110]$_{\rm tet}$ direction in the high-$T$ tetragonal notation for BaNi$_2$(As,P)$_2$.\@ Measuring the crystal's expansion along different in-plane directions is a very sensitive method for detecting a symmetry-breaking phase transition due to the `detwinning' of the sample as a consequence of the small force from the dilatometer \cite{Boehmer2015,Wang2016,Wang2018}.\@ The gradually increasing difference between the results along these two directions indicates that the tetragonal symmetry is broken for BaNi$_2$(As,P)$_2$ below $T_{\rm S,1} \approx 125$ K due to a second-order transition. In contrast to Fe122,\@ however, where the `detwinning' is along [110]$_{\rm tet}$, BaNi$_2$(As,P)$_2$ `detwins' along [100]$_{\rm tet}$.\@

The temperature-dependence of the (${\bar 3}$ ${\bar 3}$ ${\bar 6}$)$_{\rm tet}$ XRD reflection is illustrated as well as an example for all relevant reflections.\@ From crystallographic considerations \cite{footnote13} it follows that the symmetry reduction at $T_{\rm S,1}$ is most probably to the highest possible symmetry, i.e.,\@ from tetragonal to orthorhombic.\@ Together with the `detwinning' along [100]$_{\rm tet}$ observed in the TE data this allows us to conclude that the symmetry reduction is not to SG $Fmmm$ as found for Fe122 but rather to the maximal non-isomorphic orthorhombic subgroup $Immm$ \cite{footnote14}.\@  
Finally, around $T_{\rm S,2} \approx 106$ K the established first-order transition with an opposite `detwinning effect' compared to the one at $T_{\rm S,1}$ takes place (see Fig.\@ \ref{fig1}).\@ According to the strong splitting of the representative (${\bar 3}$ ${\bar 3}$ ${\bar 6}$)$_{\rm tet}$ reflection and our refinements shown in the Appendix,\@ the symmetry is at $T_{\rm S,2}$ reduced to the triclinic SG P${\bar 1}$.\@ We note that, apart from shifts in the transition temperatures, the Ni122 and the Ba(Ni,Co)$_2$As$_2$ samples show qualitatively the same behavior (see Appendices \ref{A} and \ref{B});\@ however, the second-order phase transition is most pronounced for BaNi$_2$(As,P)$_2$.\@  
\begin{table}[b]
\caption{\label{tab:table1} Values at 90 K (i.e., below $T_{\rm S,2}$) of the four different Ni-Ni bond lengths determined from refinement in $P$${\bar 1}$ for the investigated systems together with the difference $\Delta d = d_2 - d_1$.\@}
\begin{ruledtabular}
\begin{tabular}{cccc}
     & BaNi$_2$As$_2$ & BaNi$_2$(As,P)$_2$ & Ba(Ni,Co)$_2$As$_2$  \\
\hline
$d_1$ (\AA)  & 2.761(16)    & 2.800(7)      &  2.794(13)      \\
$d_2$ (\AA)  & 2.806(15)    & 2.916(7)      &  2.867(11)      \\
$d_3$ (\AA)  & 3.079(14)    & 2.942(7)      &  3.018(11)      \\
$d_4$ (\AA)  & 3.117(16)    & 3.038(7)      &  3.071(13)      \\
$\Delta d = d_2 - d_1$ (\AA) & 0.045(22)   &  0.116(11)  & 0.073(17)  \\
\end{tabular}
\end{ruledtabular}
\end{table}

With the identification of a symmetry-broken phase between the high- and low-$T$ structure, which lays the foundation for a possible nematicity, we find from our XRD data refinement in SG $Immm$ (see Appendix) for BaNi$_2$(As,P)$_2$ as well as for Ba(Ni,Co)$_2$As$_2$ that the Ni atoms move away from their special position $\frac{1}{2}$,0,$\frac{1}{4}$ in the high-$T$ phase to a new equilibrium position $\frac{1}{2}$,0,$\frac{1}{4}+\delta$ in the orthorhombic phase ($T_{\rm S,2} \leq T \leq T_{\rm S,1}$).\@ This leads for all NiAs$_4$ tetrahedra to two shortened and two elongated NiAs bond lengths of $\approx 2.34$ and $2.36$ {\AA} for BaNi$_2$(As,P)$_2$ and $\approx 2.34$ and $2.37$ {\AA} for Ba(Ni,Co)$_2$As$_2$,\@ thereby lifting the degeneracy of the $d_{xz}$ and $d_{yz}$ orbitals, as indicated in the inset of Fig.\@ \ref{fig6}(b).\@ In the absence of magnetic interactions such as super- or double-exchange, the crystal field dominates the energy levels of the orbitals and, thus, results in a staggered ordering of the $d_{xz}$ orbitals along the crystallographic $a$ axis ($d_{yz}$ orbitals along the $b$ axis for the corresponding twin) which is a further indispensable feature for charge/orbital-fluctuation-induced nematicity.
\begin{figure}[t]
\hspace{-2mm}
\includegraphics[width=0.49\textwidth]{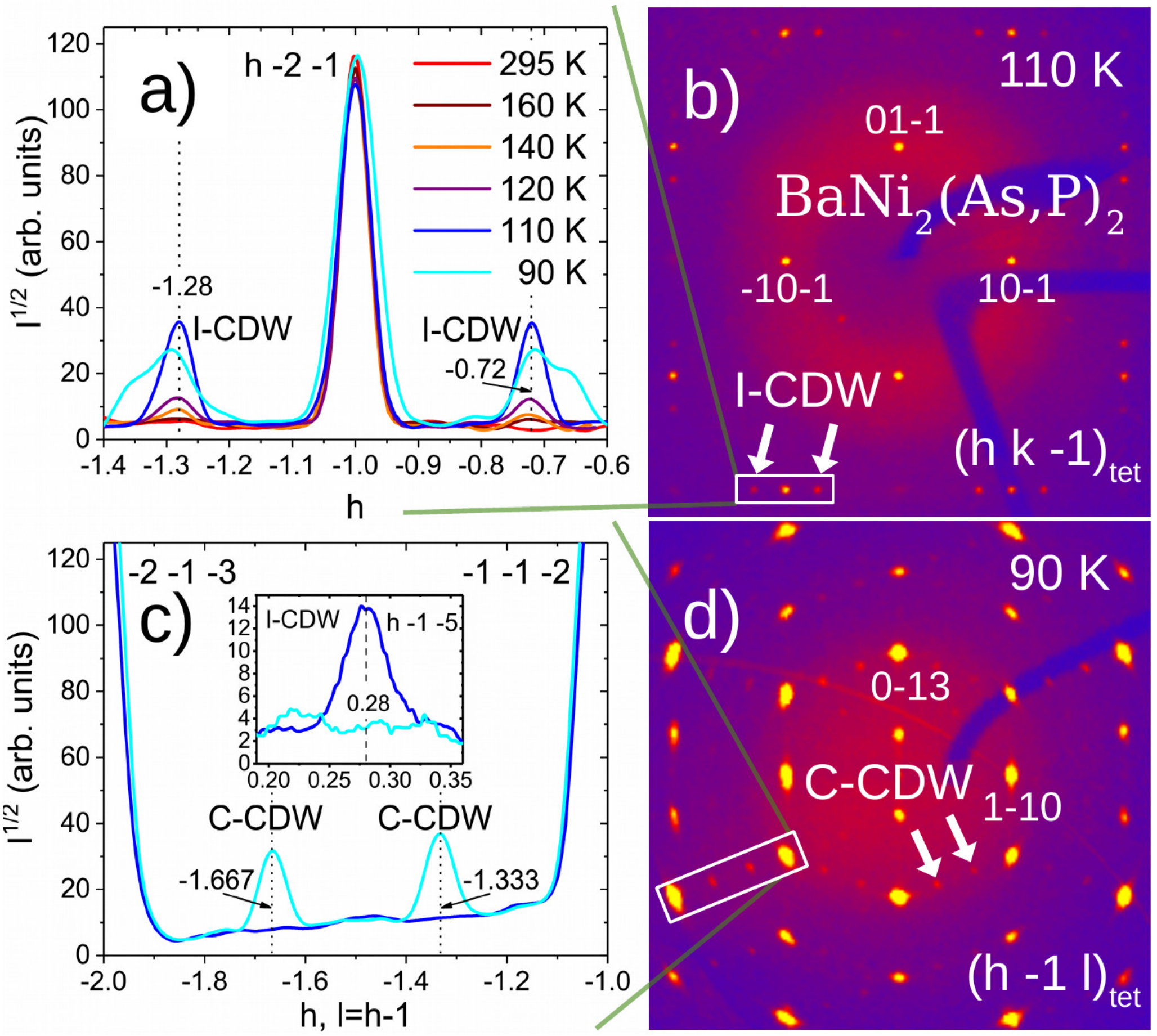}
\caption{\label{fig3}(Color online) (a) and (b): Temperature dependence of an ($h$ $k$ ${\bar 1}$)$_{\rm tet}$ plane of the RL for BaNi$_2$(As,P)$_2$.\@ Below $T_{\rm S,1} \approx 125$ K incommensurate ($h\pm0.28$ $k$ $l$)$_{\rm tet}$ and ($h$ $k\pm0.28$ $l$)$_{\rm tet}$ superstructure reflections appear around the fundamental ones as shown for (${\bar 1}$ ${\bar 2}$ ${\bar 1}$) as an example. (c) and (d): Temperature dependence of the corresponding ($h$ ${\bar 1}$ $l$)$_{\rm tet}$ plane. Below $T_{\rm S,1}$ I-CDW peaks appear (inset). Below $T_{\rm S,2}$ the I-CDW peaks disappear (inset) and commensurate ($h \pm \frac{1}{3}$ $k$ $l \mp \frac{1}{3}$)$_{\rm tet}$ superstructure reflections appear as shown for representative examples (main figure).}
\end{figure}

For completeness, we present the changes observed for the low-$T$ spatial structure. The triclinic structure of Ni122, BaNi$_2$(As,P)$_2$, and Ba(Ni,Co)$_2$As$_2$ below $T_{\rm S,2}$,\@ for which the Ni-Ni bonds approximately form zig-zag chains, is shown in Fig.\@ \ref{fig6}(c) together with an illustration of a single Ni layer with a view perpendicular to an ($0 0 1$) plane in Fig.\@ \ref{fig6}(d).\@ The four 
different Ni-Ni bond distances which are induced by the symmetry reduction to $P$${\bar 1}$ are depicted in Fig.\@ \ref{fig6}(d),\@ and the values derived from the structural refinement for the investigated systems are listed in Table \ref{tab:table1}. Full structural details of the refinement are given in the Appendix.\@ From the values listed in Table \ref{tab:table1} it is obvious that for all systems, $d_3$ and $d_4$ are much larger than $d_1$ and $d_2$.\@ It is also evident, however, that there is a certain difference between $d_1$ and $d_2$.\@ While $\Delta d = d_2 - d_1$ is relatively small for Ni122 it significantly increases for BaNi$_2$(As,P)$_2$ and Ba(Ni,Co)$_2$As$_2$.\@ This implies that the Ni-Ni zig-zag chains might still be a more or less decent approximation for Ni122,\@ but for BaNi$_2$(As,P)$_2$ and Ba(Ni,Co)$_2$As$_2$ this picture seems to be inadequate.\@ In accord with the values given in Table \ref{tab:table1} the significantly distorted zig-zag chains might be better described as Ni `dimers' connecting two Ni atoms via the short $d_1$ bond.\@ The connection of these `dimers' to the environment finally favors $d_2$ over $d_3$ and $d_4$.\@ In the context of the orbital physics discussed below, we will return to these aspects of the low-$T$ structure. The Ni-Ni dimers found here might be structural proof for the dimers resulting from orbital-selective Peierls transitions as discussed in \cite{Streltsov2014} and could explain the simultaneous appearance of both charge density wave and metallic character.

With the knowledge of a $d_{xz}$ orbital-ordered orthorhombic phase below $T_{\rm S,1}$, for which the four-fold symmetry is broken, and the structural details of the triclinic phase we will now have a closer look at the $T$-dependent development of the charge ordering probed by our XRD studies. To this end, Fig.\@ \ref{fig3}(a) and (b) present the temperature dependence of an ($h$ $k$ ${\bar 1}$)$_{\rm tet}$ plane of the reciprocal lattice (RL) for BaNi$_2$(As,P)$_2$. Despite the fact that all structure refinements have been performed in their relevant SGs, we will use the tetragonal notation throughout the paper to facilitate comparisons and to improve legibility. While larger regions of the RL are illustrated in Fig.\@ \ref{fig3}(b), we concentrate in (a) on the representative region around the (${\bar 1}$ ${\bar 2}$ ${\bar 1}$) reflection.  In the orthorhombic phase below $T_{\rm S,1} \approx 125$ K, incommensurate ($h\pm0.28$ $k$ $l$)$_{\rm tet}$ superstructure reflections and due to the twinning, ($h$ $k\pm0.28$ $l$)$_{\rm tet}$ ones appear around the fundamental reflections; they become blurry, seem to split, and the center of the split peaks moves to $\frac{1}{3}$ in the triclinic phase below $T_{\rm S,2} \approx 106$ K \cite{footnote16}.\@ Consistent with Ref.\@ \cite{Lee2019} we attribute the ($h\pm0.28$ $k$ $l$)$_{\rm tet}$  reflections to an incommensurate charge density wave (I-CDW). First indications for diffuse I-CDW peaks appear already above $T_{\rm S,1}$.\@

To complete the picture of the charge ordering for BaNi$_2$(As,P)$_2$, the temperature dependence of an ($h$ ${\bar 1}$ $l$)$_{\rm tet}$ plane is illustrated in Figs.\@ \ref{fig3}(c) and (d).\@ Weak I-CDW reflections are observed for this plane below $T_{\rm S,1}$ [see inset of Fig.\@ \ref{fig3}(c)].\@ Below $T_{\rm S,2}$, however, the I-CDW peaks disappear and, as show in Figs.\@ \ref{fig3}(c) and (d),\@ commensurate ($h \pm \frac{1}{3}$ $k$ $l \mp \frac{1}{3}$)$_{\rm tet}$ reflections appear around the fundamental reflections (together with the reflections of the twins). Consistent with Ref.\@ \cite{Lee2019} we attribute the ($h \pm \frac{1}{3}$ $k$ $l \mp \frac{1}{3}$)$_{\rm tet}$ reflections to the C-CDW expected for the ground state of charge/orbital-fluctuation-induced nematicity as outlined above.\@ Corresponding effects with identical {\bf q} vectors of the I-CDW and of the C-CDW, respectively, but with different values for $T_{\rm S,1}$ and $T_{\rm S,2}$ are found for Ni122 and Ba(Ni,Co)$_2$As$_2$ as well (see Appendices \ref{A} and \ref{B}).\@ The 'splitting' of the I-CDW reflections observed for the low-$T$ phase in Fig.\@ \ref{fig3}(a) together with the movement from $h = 0.28$ to $\frac{1}{3}$ and the vanishing of the I-CDW peaks in Fig.\@ \ref{fig3}(c) suggests that the ground-state C-CDW develops directly from the I-CDW.\@

\begin{figure}[t]
\includegraphics[width=0.4925\textwidth]{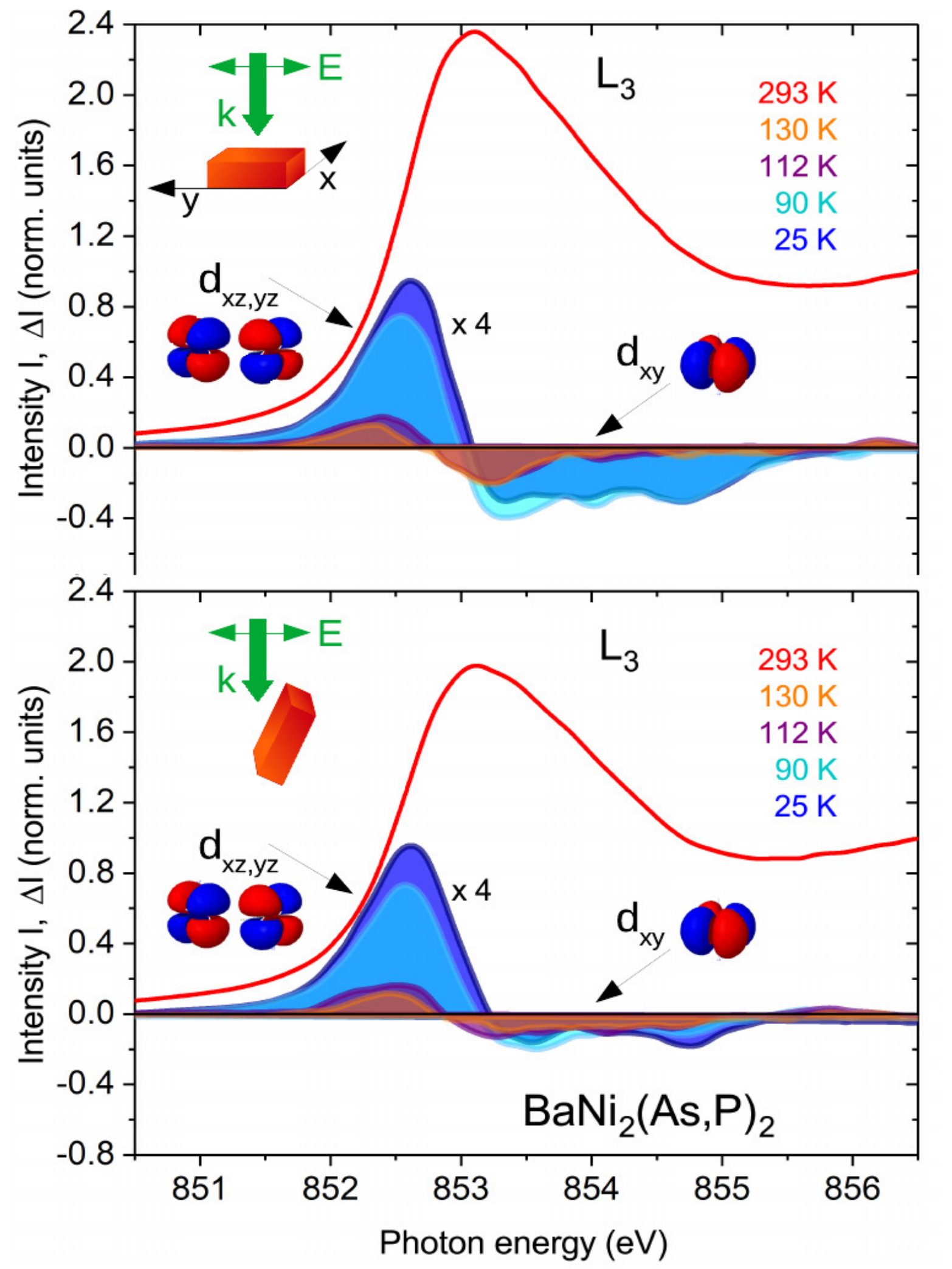}
\caption{\label{fig5p}(Color online) Normal and grazing incidence NEXAFS of BaNi$_2$(As,P)$_2$.\@ The sketch in the upper left corner illustrates the polarization of light with respect to the sample's orientation. Shown is the spectrum taken at 293 K and the difference $\Delta I = I$(293 K)$-I$($T$) between the spectra taken at 293 K and at the respective temperatures given in the graph. Please note that $\Delta I$ is multiplied by a factor of 4 and that a positive area means \textit{loss} of spectral weight relative to RT.\@ A significant charge transfer from orbitals with in- and out-of-plane character, i.e., $d_{xz,yz}$ orbitals, to orbitals with exclusively in-plane character, i.e., to $d_{xy}$ orbitals is found below $T_{\rm S,2} \approx 106$ K. First indications for this transfer are, however, observed already above $T_{\rm S,1}$.\@}
\end{figure}
Finally, we want to see if and how the charge-ordering effects are tied to the orbital physics at the Ni sites.\@ NEXAFS is uniquely suited for this since the excitation process is element-specific (Ni $2p \rightarrow 3d$) as well as polarization- and, thus, orbital-dependent. 
In the present context, the $e$-type orbitals ($d_{x^2 - y^2}$ and $d_{3z^2 - r^2}$) are essentially filled \cite{Merz2016} and spectroscopically silent. Hence Ni $L$-edge NEXAFS probes just the $t_2$ orbitals here: the $d_{xz,yz}$ orbitals appear equally in both normal and grazing incidence, while the $d_{xy}$ orbitals appear predominantly in normal incidence (cf.\@ also Figs.\@ \ref{fig6}d and \ref{fig5p}).\@ For BaNi$_2$(As,P)$_2$, Fig.\@ \ref{fig5p} depicts the spectral evolution with temperature --- first the room-temperature curve I(293 K) \cite{footnote15}, and then, as shaded areas for several decreasing values of $T$, the changes relative to it, $\Delta I = I$(293 K)$-I$($T$)\@. The distinct behavior for normal and grazing incidence enables the identification of the orbitals involved and clearly demonstrates that for temperatures below $T_{\rm S,2} \approx 106$ K,\@ charge carriers are increasingly transferred from $d_{xz,yz}$ orbitals (at 'low' energy) to $d_{xy}$ orbitals (at 'higher' energy). This transfer of charge carriers towards $d_{xy}$ with decreasing $T$ is most pronounced for Ni122 and less so for BaNi$_2$(As,P)$_2$ and Ba(Ni,Co)$_2$As$_2$, cf. Fig.\@ \ref{fig5p} and Appendices \ref{A} and \ref{B}, indicating different levels of importance of the $d_{xy}$ orbitals in the formation of the Ni-related geometry. And indeed, the picture is entirely consistent to our XRD findings: for Ni122, the Ni-Ni zigzag chains (and the C-CDW) emerge whose formation requires a prominent contribution from $d_{xy}$;\@ and for BaNi$_2$(As,P)$_2$ and Ba(Ni,Co)$_2$As$_2$,\@ we have the connected Ni-Ni 'dimers' for which the $d_{xy}$ orbitals play a less important role than the $d_{xz,yz}$ orbitals. Interestingly, a small amount of charge carriers is transferred between $d_{xz,yz}$ and $d_{xy}$ orbitals already above $T_{\rm S,1}$,\@ i. e., above the orthorhombic phase transition where the signatures of the I-CDW peaks have a diffuse character as discussed above. Both findings point to dynamic effects and, thus, to charge/orbital fluctuations in this $T$ range (see also the NEXAFS spectra of Ni122 and Ba(Ni,Co)$_2$As$_2$ in Appendices \ref{A} and \ref{B} where indications for charge/orbital fluctuations are found already far above the nematic/structural transition at $T_{\rm S,1}$).\@ As discussed above, charge fluctuations of this type are an essential prerequisite for charge/orbital induced nematicity. For Fe122, on the other hand, such a charge transfer is completely absent in the whole investigated temperature range (see Appendix \ref{AppC}), ruling out this kind of charge/orbital induced nematicity there.

\section{Summary and Conclusions}

Taken together, the results presented here demonstrate that Ba(Ni,Co)$_2$(As,P)$_2$ is a promising candidate for a system with charge/orbital-fluctuation-induced nematicity.\@ All necessary prerequisites for charge/orbital-fluctuation-induced nematicity are met such as charge/orbital fluctuations above $T_{\rm nem}$ ($= T_{\rm S,1}$ in our measurements),\@ a second-order nematic/structural transition for which the fourfold symmetry breaking is accompanied by orbital order in the temperature range $T_{\rm CDW} \leq T \leq T_{\rm nem}$, and finally a first order transition for which the translation symmetry is broken in the CDW ground state below $T_{\rm CDW}$ ($= T_{\rm S,2}$ in our measurements).\@ Another argument supporting charge/orbital fluctuations might be that the $d_{xz}$ orbitals continously order as staggered chains along [100]$_{\rm tet}$ in the nematic phase of Ba(Ni,Co)$_2$(As,P)$_2$ in contrast to the ordering along [110]$_{\rm tet}$ for Fe122 where it is assumed that the orbital order is induced by magnetic fluctuations. Furthermore, it is sometimes proposed that the pnictogen height, $h$,\@ plays an important role for the effective bandwidth of magnetic excitations \cite{Zhang2014}.\@ Our more recent studies show that, for example, FeSe with $h = 1.473$ {\AA} at room temperature has a $T_{\rm c} \approx 9$ K whereas Fe$_{1+\delta}$Te with a strongly increased $h = 1.764$ {\AA} has an antiferromagnetic low-$T$ phase. In this context, the significantly reduced $h$ when going from Fe122 ($h = 1.355$ {\AA}) to Ni122 ($h = 1.134$ {\AA}) \cite{Merz2016} might prevent Ni122 from developing magnetic fluctuations, let alone antiferromagnetism.\@ 

We conclude with some findings for the relation between the ground state CDW phase and $T_{\rm c}$:\@ According to our data, the $d_{xy}$ orbitals are very important for the zig-zag chains of Ni122 while they play a lesser role for the connected Ni-Ni 'dimers' of BaNi$_2$(As,P)$_2$,\@ and Ba(Ni,Co)$_2$As$_2$ where the contribution of the $d_{xy}$ ($d_{xz,yz}$) orbitals is gradually reduced (enhanced) with increasing substitution.\@ Taking into account the postulated phase diagrams of Ref. \cite{Kudo2012} and \cite{Lee2019},\@ the boost of $T_{\rm c}$ to $\sim 3.6$ K sets finally in via a first-order transition when the distorted zig-zag chains disappear together with the C-CDW at the structural phase boundary. In this sense, long-range charge ordering effects involving $d_{xy}$ orbitals seem to be detrimental to superconductivity in Ba(Ni,Co)$_2$(As,P)$_2$.\@ Our data suggest that a delicate balance of the $d_{xz,yz}$ and $d_{xy}$ orbital-dependent density of states decides whether long-range charge/orbital order or enhanced superconductivity is formed.

\begin{acknowledgments}
We are indebted to A.\@ Haghighirad,\@  R.\@ Heid,\@  J.\@ Schmalian,\@  
S.\@  M.\@  Souliou,\@  and M.\@ Le Tacon for fruitful discussions. 
We gratefully acknowledge the Synchrotron Light Source KARA Karlsruhe and the Karlsruhe Nano Micro Facility for Information (KNMFi) for the provision of beamtime.\@ L.\@ W.\@ thanks the support from Deutsche Forschungsgemeinschaft (DFG) through Grant No.WA4313/1-1.\@
\end{acknowledgments}

\appendix

\begin{figure*}[t]
\hspace{-3mm}
\includegraphics[width=0.475\textwidth]{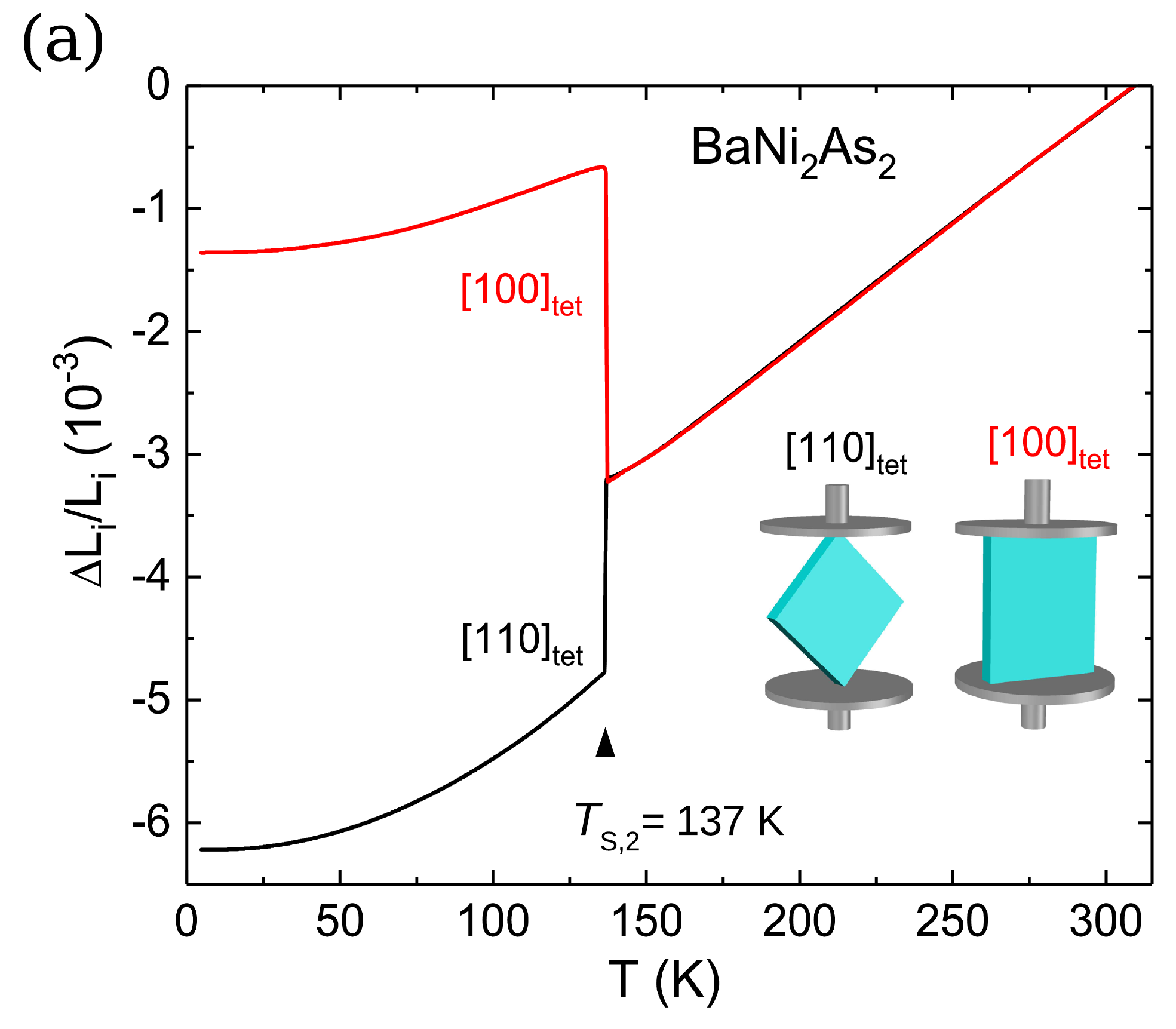}
\hspace{3mm}
\includegraphics[width=0.475\textwidth]{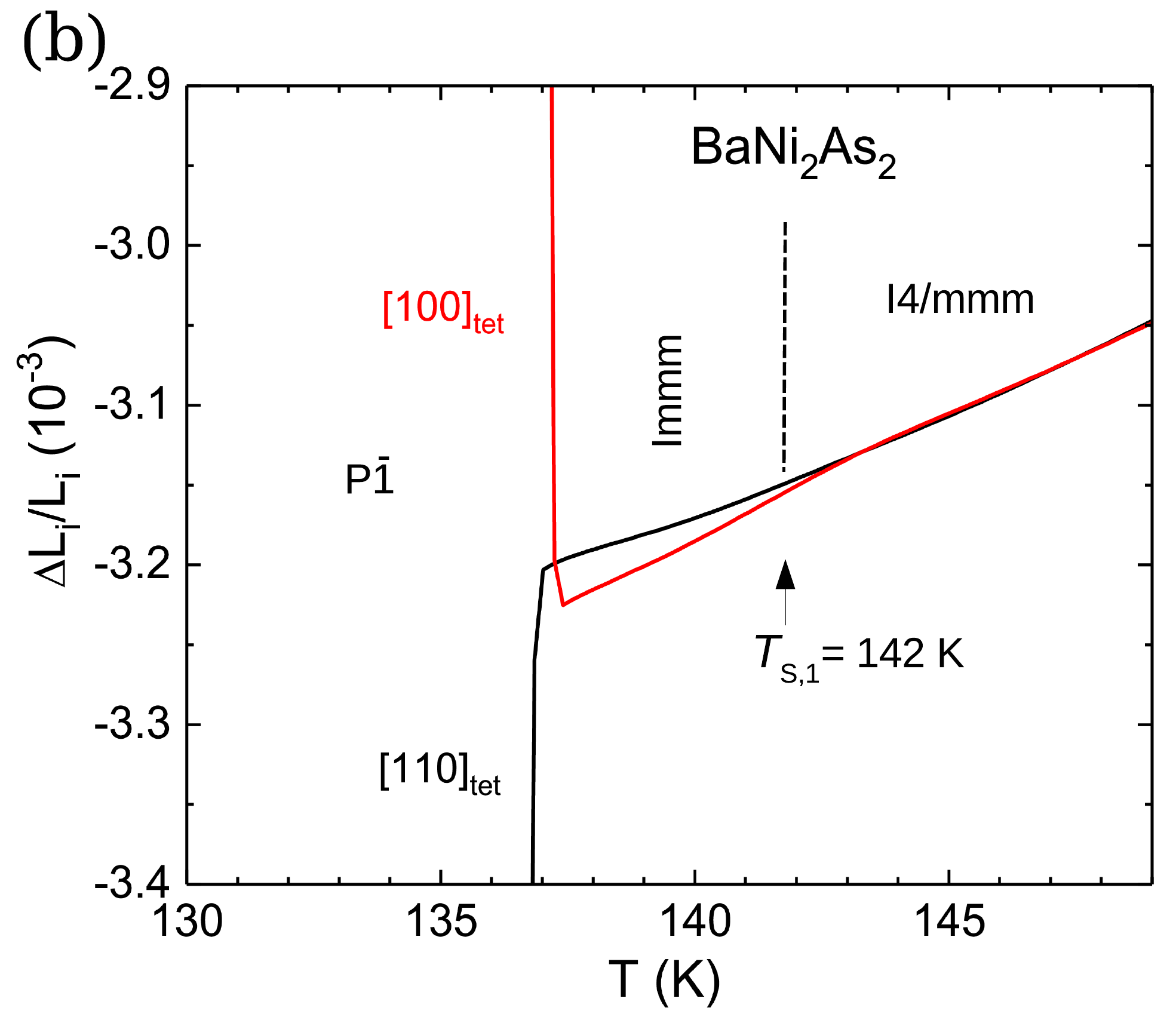}
\caption{\label{figS1} (Color online) (a) Relative length change, $\Delta L/L$, versus temperature of the [100]$_{\rm tet}$ and the [110]$_{\rm tet}$ direction in the high-$T$ tetragonal notation for BaNi$_2$As$_2$ obtained using high-resolution capacitance dilatometry and upon heating. The whole measured temperature range is plotted in the figure. (b) $\Delta L/L$ for BaNi$_2$As$_2$ displayed in an enlarged section of the relevant range above the first order transition which shows the gradually increasing difference between the results along the two directions, and, thus, the symmetry breaking.}
\end{figure*}

\section{\label{A} Measurements on BaNi$_2$As$_2$}

In the following we will present the complementary results measured with TE, XRD, and NEXAFS on BaNi$_2$As$_2$ (Ni122), Ba(Ni,Co)$_2$As$_2$,\@ and BaFe$_2$As$_2$ (Fe122),\@ respectively. Furthermore, we will show Tables which summarize our temperature-dependent crystallographic data.

\subsection{Thermal expansion of BaNi$_2$As$_2$}

Fig.\@ \ref{figS1}(a) presents the high-resolution thermal expansion, $\Delta L/L$,\@ of Ni122 measured along the [100]$_{\rm tet}$ and the [110]$_{\rm tet}$ direction in the high-$T$ tetragonal notation.\@ The orientation of the sample is sketched in the inset of Fig.\@ \ref{figS1}(a).\@ A pronounced first-order transition is observed below $T_{\rm S,2} \approx 137$ K.\@  Compared to BaNi$_2$(As,P)$_2$ (see Fig.\@ 1) an even stronger difference between the [100]$_{\rm tet}$ and the [110]$_{\rm tet}$ direction is found for Ni122 and $T_{\rm S,2}$ is increased by $\approx 30$ K.\@

In contrast to BaNi$_2$(As,P)$_2$,\@ however, the second order transition is weaker for Ni122.\@ To show the gradually increasing difference between the results along the two directions upon cooling, an enlarged section of the relevant temperature range above the first order transition is displayed in Fig.\@ \ref{figS1}(b).\@ Even though this difference is very small, the symmetry breaking below $T_{\rm S,1} \approx 142$ K is evident from our TE data on Ni122.

\subsection{X-ray diffraction on BaNi$_2$As$_2$}

\subsubsection{Temperature dependence of the (${\bar 3}$ ${\bar 3}$ ${\bar 6}$)$_{\rm tet}$ fundamental reflection}

\begin{figure}[b]
\includegraphics[width=0.475\textwidth]{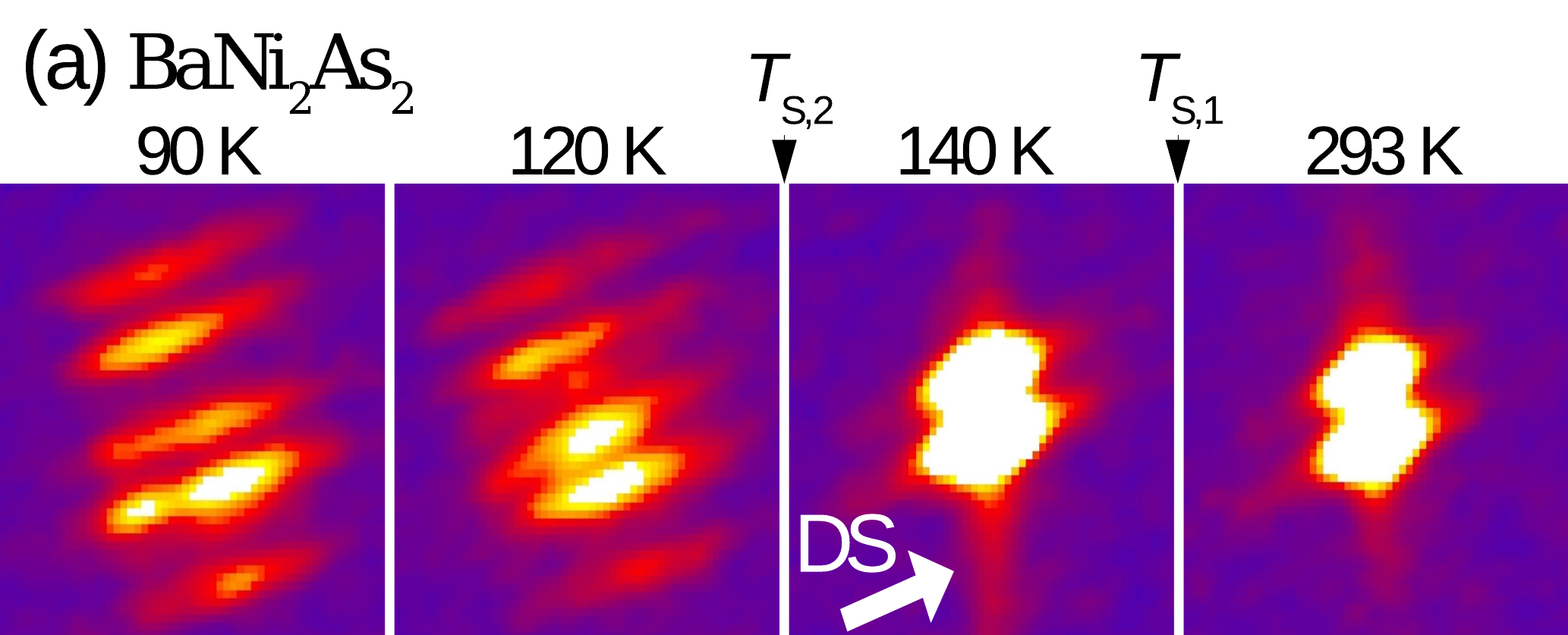}\\[2mm]
\includegraphics[width=0.475\textwidth]{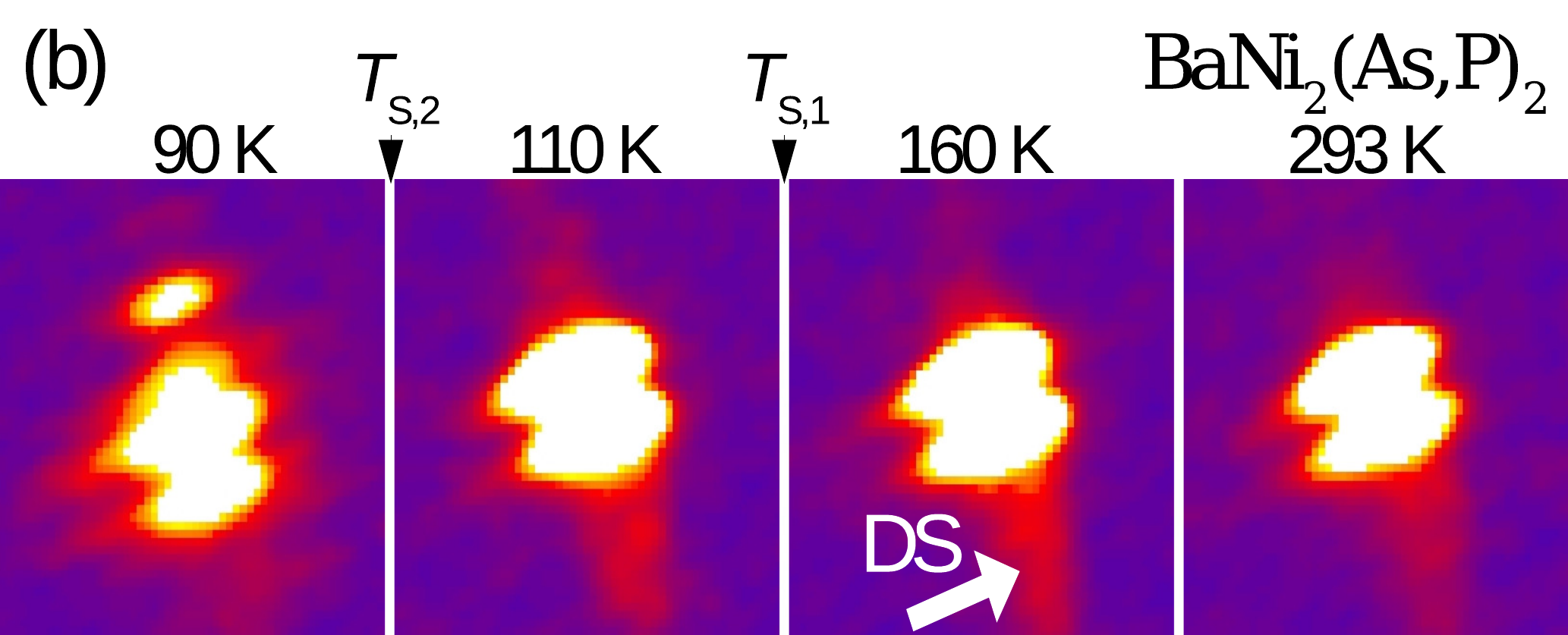}
\caption{\label{figS2} (Color online) Splitting of the (${\bar 3}$ ${\bar 3}$ ${\bar 6}$)$_{\rm tet}$ reflection at the structural phase transition of (a) BaNi$_2$As$_2$,\@ and (b) BaNi$_2$(As,P)$_2$.\@ For both samples the structural transition to triclinic is clearly resolved by the peak splitting below $T_{\rm S,2}$.\@ A significant increase in diffuse scattering is observed upon cooling.} 
\end{figure}
In Fig.\@ \ref{figS2}(a) the temperature dependence of the (${\bar 3}$ ${\bar 3}$ ${\bar 6}$)$_{\rm tet}$ XRD reflection of Ni122 is illustrated as a representative for all relevant reflections. In contrast to the TE data in Fig.\@ \ref{figS1}(b),\@ a prominent signature for a symmetry breaking such as a clear splitting of the fundamental reflections, is absent in the temperature range $T_{\rm S,2} \leq T \leq T_{\rm S,1}$,\@ i.e.\@ in the orthorhombic phase, which we attribute to the fact that the change of the lattice parameters induced by the symmetry breaking is very small and that our high-resolution capacitance dilatometer is several orders of magnitude more sensitive than standard diffraction techniques. What can be detected, however, is that the (${\bar 3}$ ${\bar 3}$ ${\bar 6}$)$_{\rm tet}$ reflection broadens slightly below $T_{\rm S,1} = 142$ K.\@ Within the resolution of the experiment and taking the tiny change of the lattice parameters found in the TE experiment [see Fig.\@ \ref{figS1}(b)] into account, such a broadening is a good indicator for a symmetry reduction at the orthorhombic/nematic transition below $T_{\rm S,1}$.\@ Consistent with the change of lattice parameters found in the TE experiment [shown in Fig.\@ 1], a similar $T$-dependent broadening effect of the (${\bar 3}$ ${\bar 3}$ ${\bar 6}$)$_{\rm tet}$ reflection is found for BaNi$_2$(As,P)$_2$ as well [see Fig.\@ \ref{figS2}(b)[ which corroborates the symmetry reduction from tetragonal to orthorhombic below $T_{\rm S,1}$ also for this compound.

In addition to the peak broadening a general increase in diffuse scattering is observed upon cooling in the immediate vicinity of the fundamental reflections which grows already above $T_{\rm S,1}$.\@ This increase in diffuse scattering (DS) displayed for Ni122 and BaNi$_2$(As,P)$_2$ in Figs.\@ \ref{figS2}(a) and (b), respectively, 
cannot originate from thermal diffuse scattering (TDS) since TDS usually decreases upon cooling.\@ The typical behavior of TDS is depicted for Fe122 in Fig.\@ \ref{figS10} for comparison. Furthermore, it is not expected that disorder increases upon cooling and, therefore, the observed DS could be induced by charge/orbital fluctuations and supports a picture where the tetragonal to orthorhombic transition can be interpreted as nematic. In this sense the significant increase of DS demonstrates that the charge/orbital fluctuations strongly interact with the lattice which might be reflected in future Raman investigations.

Below $T_{\rm S,2} = 137$ K a very strong splitting of the (${\bar 3}$ ${\bar 3}$ ${\bar 6}$)$_{\rm tet}$ reflection of BaNi$_2$As$_2$ is observed in Fig.\@ \ref{figS2}(a) which is consistent with the symmetry reduction to triclinic. In line with our TE results it can easily be seen that the triclinic distortion is much more pronounced for BaNi$_2$As$_2$ [see Fig.\@ \ref{figS2}(a)] than for BaNi$_2$(As,P)$_2$ [see Fig.\@ \ref{figS2}(b)] where $T_{\rm S,2} = 106$ K.

\subsubsection{Temperature dependence of the two types of charge density waves}

In Fig.\@ 3 we have shown that two types of charge density waves are observed for BaNi$_2$(As,P)$_2$ upon cooling: (i) an incommensurate charge density wave (I-CDW) where ($h \pm 0.28$ $k$ $l$)$_{\rm tet}$ superstructure reflections appear around the fundamental reflections in the orthorhombic phase and (ii) a commensurate charge density wave (C-CDW) where ($h \pm \frac{1}{3}$ $k$ $l \mp \frac{1}{3}$)$_{\rm tet}$ superstructure reflections appear in the triclinic phase. As described above we use the tetragonal notation to facilitate comparisons and to improve legibility. In the following we will discuss the temperature dependence of the two types of charge density waves for Ni122.\@

\begin{figure}[t]
\hspace{-0mm}
\includegraphics[width=0.49\textwidth]{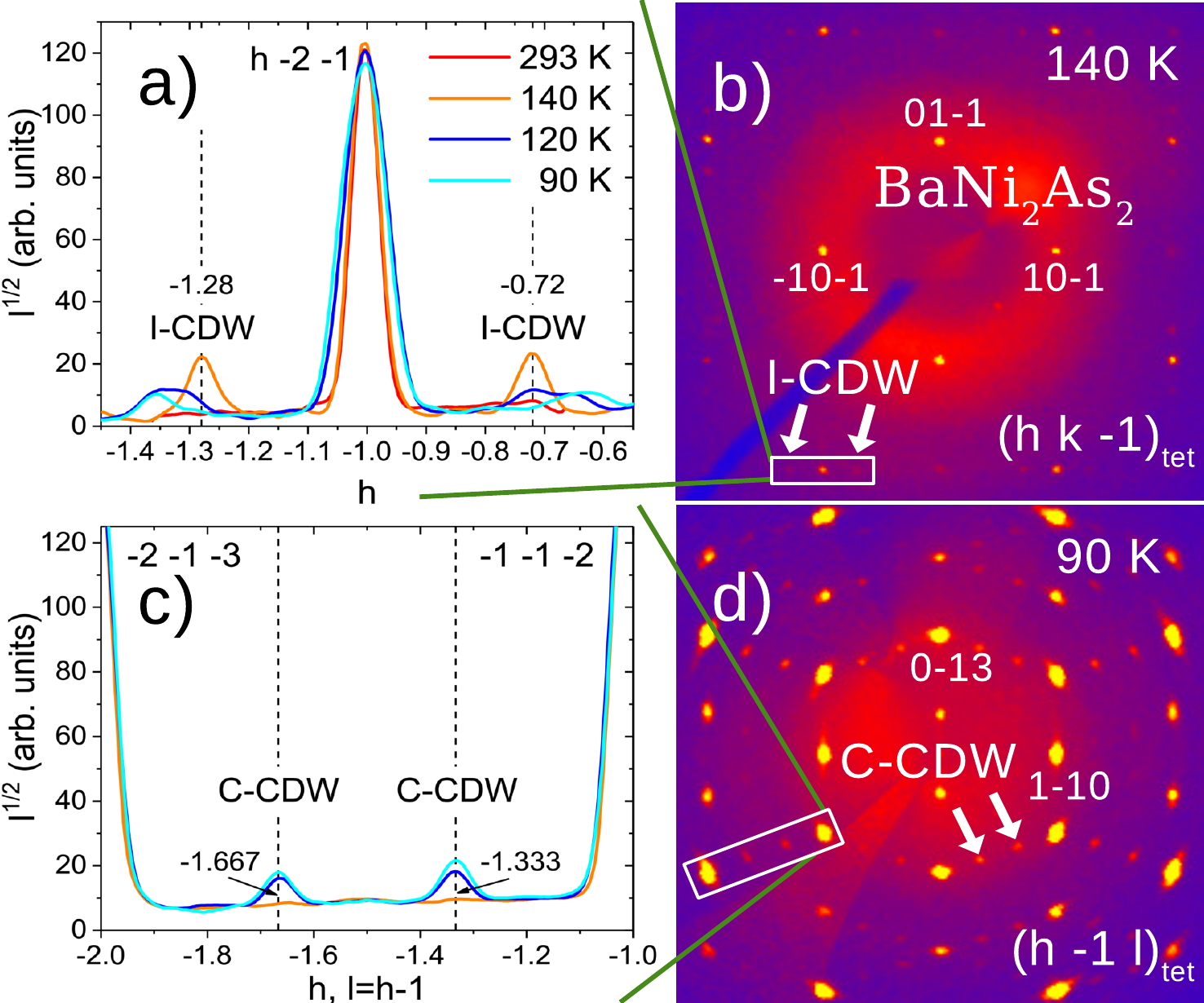}\\
\caption{\label{figS3}(Color online) (a) and (b) Temperature dependence of an ($h$ $k$ ${\bar 1}$)$_{\rm tet}$ plane of the reciprocal lattice for BaNi$_2$As$_2$. Below $T_{\rm S,1} \approx 142$ K incommensurate ($h \pm 0.28$ $k$ $l$)$_{\rm tet}$ and ($h$ $k \pm 0.28$ $l$)$_{\rm tet}$ superstructure reflections appear around the fundamental reflections and become blurry and seem to split below $T_{\rm S,2} \approx 137$ K. (c) and (d) Temperature dependence of an ($h$ ${\bar 1}$ $l$)$_{\rm tet}$ plane of the reciprocal lattice for BaNi$_2$As$_2$. Below $T_{\rm S,2} \approx 137$ K commensurate ($h \pm \frac{1}{3}$ $k$ $l \mp \frac{1}{3}$)$_{\rm tet}$ superstructure reflections appear around the fundamental reflections. In (b) and (d) arrows mark the positions of representative superstructure reflections.}
\end{figure}

To this end, Fig.\@ \ref{figS3}(a) and (b) present the temperature dependence of an ($h$ $k$ ${\bar 1}$)$_{\rm tet}$ plane of the reciprocal lattice for Ni122. In the orthorhombic phase below $T_{\rm S,1} \approx 142$ K, incommensurate ($h \pm 0.28$ $k$ $l$)$_{\rm tet}$  superstructure reflections, and due to the twinning, ($h$ $k \pm 0.28$ $l$)$_{\rm tet}$ ones appear around the fundamental reflections. In contrast to BaNi$_2$(As,P)$_2$,\@ however, the incommensurate ($h \pm 0.28$ $k$ $l$)$_{\rm tet}$ superstructure reflections are significantly weaker for Ni122.\@ In the triclinic phase below $T_{\rm S,2} \approx 137$ K the superstructure reflections smear out, seem to 'split', and the center of the split peaks moves from $h = 0.28$ towards $\frac{1}{3}$ \cite{footnote16}.\@ Upon further cooling one of the 'split' C-CDW reflections intensifies at the expense of the other.

To complete the picture of the charge ordering for BaNi$_2$As$_2$, the temperature dependence of an ($h$ ${\bar 1}$ $l$)$_{\rm tet}$ plane is illustrated in Fig.\@ \ref{figS3}(c) and (d).\@ Very weak I-CDW reflections are observed below $T_{\rm S,1}$ for the ($h$ ${\bar 2}$ $l$)$_{\rm tet}$ plane (not shown).\@ Below $T_{\rm S,2}$, however, the weak I-CDW peaks disappear and commensurate ($h \pm \frac{1}{3}$ $k$ $l \mp \frac{1}{3}$)$_{\rm tet}$ superstructure reflections appear around the fundamental reflections. Consistent with Ref.\@ \cite{Lee2019} we attribute the ($h \pm \frac{1}{3}$ $k$ $l \mp \frac{1}{3}$)$_{\rm tet}$ reflections to a commensurate charge density wave (C-CDW). In analogy to the discussion of BaNi$_2$(As,P)$_2$,\@ we conclude that the C-CDW develops directly from the I-CDW.\@ By comparing the temperature dependence of the two types of CDWs for Ni122 and BaNi$_2$(As,P)$_2$ it seems to be evident that the intensity of the I-CDW normalized to their corresponding fundamental reflections is directly related to the orthorhombic distortion.\@ The normalized intensity of the C-CDW, however, does obviously not scale with the triclinic distortion but is rather correlated with the I-CDW observed in the orthorhombic phase (see also discussion in Appendix \ref{AppA2c}).

\subsubsection{\label{AppA2c}XRD: Structure determination and refinement}

Using Mo $K_{\alpha}$ radiation all accessible symmetry-equivalent reflections were measured up to a maximum angle of $2 \Theta =65^{\circ}$.\@ The data were corrected for Lorentz, polarization, extinction, and absorption effects.\@ Using SHELXL \cite{Sheldrick} and JANA2006 \cite{Petricek}, all averaged symmetry-independent reflections ($I > 2 \sigma$) have been included for the respective refinements in their corresponding space groups. For all temperature ranges the unit cell and the space group were determined, the atoms were localized in the unit cell utilizing random phases as well as Patterson superposition methods, the structure was completed and solved using difference Fourier analysis, and finally the structure was refined. For the orthorhombic and the triclinic phases the corresponding twinning was taken into account. All refinements converged quite well and show excellent reliability factors ($wR_2$, $R_1$, GOF).\@ 
\begin{table}[t]
\caption{\label{tab:table1} Crystallographic data for BaNi$_2$As$_2$ at 295 K,\@ 140 K,\@ and 90 K determined from single-crystal x-ray diffraction. The high-temperature structure was refined in the tetragonal space group (SG) $I4/mmm$,\@ the nematic phase was refined in the orthorhombic SG $Immm$,\@ and the low-temperature structure was refined in the triclinic SG $P$${\bar 1}$.\@ $U_{\rm iso}$ denotes the isotropic atomic displacement parameters (ADP). The ADPs were refined anisotropically but due to space limitations only the $U_{\rm iso}$ are listed in the table. The Wyckoff positions (Wyck.) are given for their respective space groups. TW represents the degree of twinning in the corresponding phases. Errors shown are statistical errors from the refinement.}
\begin{ruledtabular}
		\begin{tabular}[t]{llccc}
			%\captionof{table}{bla}
            &                           &       295 K          &          140 K      &       90 K \\ \hline
			& SG                        &     $I4/mmm$         &      $Immm$         & $P$${\bar 1}$    \\
			&  $a$ (\AA)              &      4.1413(6)       &        4.1254(7)    &       4.1508(13) \\
			&  $b$ (\AA)              &      4.1413(6)       &        4.1295(10)    &       4.1539(13) \\
			&  $c$ (\AA)              &      11.6424(24)     &       11.6463(26)   &       6.4579(22) \\
			&  $\alpha$ (\textdegree)             &       90             &          90         &      108.688(26)   \\
			&  $\beta$ (\textdegree)              &       90             &          90         &      108.650(26)   \\
			&  $\gamma$ (\textdegree)             &       90             &          90         &       90.022(26) \\
			&  V (\AA$^3$)            &      199.7           &         198.4       &       99.3 \\ %%\hline
    Ba      & Wyck.                     &      $2a$            &         $2a$        &       $1a$ \\
			& $x$                       &        0             &          0          &         0 \\
			& $y$                       &        0             &          0          &         0 \\
			& $z$                       &        0             &          0          &         0 \\
			& $U_{\rm iso}$ (\AA$^2$) &      0.01505(20)     &      0.00942(23)   &     0.02369(122) \\ %%\hline
    Ni      & Wyck.                     &       $4d$           &         $4j$        &     $2i$  \\
            & $x$                       &        $\frac{1}{2}$         &        $\frac{1}{2}$        &      0.72064(108) \\
			& $y$                       &          0           &           0         &     0.24534(249) \\
			& $z$                       &        $\frac{1}{4}$         &        0.24998(13)     &   0.49955(192) \\
			& $U_{\rm iso}$ (\AA$^2$) &  0.02143(32)         &      0.01442(36)    &      0.04369(109) \\
    As      & Wyck.                     &      $4e$            &         $4i$        &       $2i$ \\
            & $x$                       &        0             &          0          &      0.67837(71)\\
			& $y$                       &        0             &          0          &      0.65326(195) \\
			& $z$                       &    0.34727(10)        &      0.34755(9)    &      0.31105(34) \\
			& $U_{\rm iso}$ (\AA$^2$) &    0.01905(26)       &      0.01266(30)       &      0.03550(154) \\
			& TW (\%)                   &      $-$                &         57/43       &      31/30/22/17 \\
			& GOF                       &       2.33           &         2.11        &       2.80 \\
			& $wR_2$ (\%)               &       5.07           &         5.17        &       8.58  \\
			& $R_1$ (\%)                &       1.94           &         2.10        &       4.41  \\
		\end{tabular}
\end{ruledtabular}
\end{table}
\begin{table}[t]
\caption{\label{tab:table2} Crystallographic data for BaNi$_2$(As,P)$_2$ at 295 K,\@ 110 K,\@ and 90 K determined from single-crystal x-ray diffraction. The P content was determined to 3 \%.\@ The high-temperature structure was refined in the tetragonal space group (SG) $I4/mmm$,\@ the nematic phase was refined in the orthorhombic SG $Immm$,\@ and the low-temperature structure was refined in the triclinic SG $P$${\bar 1}$.\@ $U_{\rm iso}$ denotes the isotropic atomic displacement parameters (ADP). The ADPs were refined anisotropically but due to space limitations only the $U_{\rm iso}$ are listed in the table. The Wyckoff positions (Wyck.) are given for their respective space groups. TW represents the degree of twinning in the corresponding phases.}
\begin{ruledtabular}
		\begin{tabular}[t]{llccc}
            &                           &       295 K          &          110 K      &       90 K \\ \hline
			& SG                        &     $I4/mmm$         &      $Immm$         & $P$${\bar 1}$    \\
			&  $a$ (\AA)              &      4.1358(8)       &        4.1218(4)    &       4.1296(14) \\
			&  $b$ (\AA)              &      4.1358(8)       &        4.1242(4)    &       4.1375(14) \\
			&  $c$ (\AA)              &      11.6468(35)     &       11.6363(17)   &       6.4817(24) \\
			&  $\alpha$ (\textdegree)             &       90             &          90         &      108.512(27)   \\
			&  $\beta$ (\textdegree)              &       90             &          90         &      108.415(27)   \\
			&  $\gamma$ (\textdegree)             &       90             &          90         &       90.145(27) \\
			&  V (\AA$^3$)            &      199.2           &         197.8       &       99.0 \\ %%\hline
    Ba      & Wyck.                     &      $2a$            &         $2a$        &       $1a$ \\
			& $x$                       &        0             &          0          &         0 \\
			& $y$                       &        0             &          0          &         0 \\
			& $z$                       &        0             &          0          &         0 \\
			& $U_{\rm iso}$ (\AA$^2$) &      0.01337(18)     &      0.01018(22)   &     0.01773(62) \\ %%\hline
    Ni      & Wyck.                     &       $4d$           &         $4j$        &     $2i$  \\
            & $x$                       &        $\frac{1}{2}$         &        $\frac{1}{2}$        &      0.76375(77) \\
			& $y$                       &          0           &           0         &     0.25968(78) \\
			& $z$                       &        $\frac{1}{4}$         &        0.24803(15)     &      0.50058(43) \\
			& $U_{\rm iso}$ (\AA$^2$) &  0.02029(25)         &      0.01757(25)    &      0.02843(109) \\
    As/P    & Wyck.                     &      $4e$            &         $4i$        &       $2i$ \\
            & $x$                       &        0             &          0          &      0.64531(64)\\
			& $y$                       &        0             &          0          &      0.66750(65) \\
			& $z$                       &    0.34704(9)        &      0.34708(16)    &      0.30580(29) \\
			& $U_{\rm iso}$ (\AA$^2$) &    0.01738(26)       &      0.01521(28)       &      0.02515(80) \\
			& TW (\%)                   &      $-$                &         55/45       &      51/44/4/1 \\
			& GOF                       &       2.17           &         2.98        &       2.94 \\
			& $wR_2$ (\%)               &       4.59           &         6.43        &       7.89  \\
			& $R_1$ (\%)                &       1.83           &         2.44        &       3.92  \\
		\end{tabular}
\end{ruledtabular}
\end{table}

Structural details of Ni122 resulting from refinements of diffraction data measured at temperatures characteristic for the tetragonal, orthorhombic/nematic, and triclinic phase, respectively, are listed in Table \ref{tab:table1}.\@ Corresponding results for BaNi$_2$(As,P)$_2$ are compiled in Table \ref{tab:table2} for comparison. Some consequences of the structural arrangement derived from the crystallographic data shown in Table \ref{tab:table1} and \ref{tab:table2},\@ such as the (distorted) zig-zag chains of the low-$T$ triclinic structure or the staggered chains along [100]$_{\rm tet}$ in the orthorhombic/nematic structure, have already been discussed in detail above. Since a complete crystallographic discussion is beyond the scope of an Appendix we will take a closer look only on the changes of the bond distances at the phase transitions.\@ In the tetragonal phase we have one Ni-Ni and one Ni-As bond distance of 2.928 and 2.360 \AA,\@ respectively, for Ni122 and of 2.924 and 2.356 \AA,\@ respectively, for BaNi$_2$(As,P)$_2$.\@ In the orthorhombic phase we still have one Ni-Ni bond length of 2.919 {\AA} for Ni122 and of 2.916 {\AA} for BaNi$_2$(As,P)$_2$;\@ the Ni-As bond length, however, splits up: 2.355 and 2.357 {\AA} for Ni122 and 2.340 and 2.362 {\AA} for BaNi$_2$(As,P)$_2$.\@ Thus, consistently with our TE data we find a tiny (in principle, almost insignificant) orthorhombic distortion in the case of Ni122 while a significant one is observed for BaNi$_2$(As,P)$_2$.\@ In the absence of magnetic interactions, this splitting is responsible for lifting the degeneracy of the $d_{xz}$ and $d_{yz}$ orbitals and, thus, for the staggered ordering of the $d_{xz}$ orbitals along the crystallographic $a$ axis (see also Fig.\@ \ref{fig6}).\@ Obviously, the degree of orthorhombic distortion reflected by the magnitude of the Ni-As bond length splitting is directly correlated with the intensity of the I-CDW superstructure reflections:\@ Strong reflections are observed for the I-CDW of BaNi$_2$(As,P)$_2$ in Figs.\@ 3(a) and (b) while only very weak ones are detected for Ni122 in Figs.\@ \ref{figS3}(a) and (b).\@ This finding suggests that the staggered ordering of the $d_{xz}$ orbitals in the orthorhombic/nematic phase plays a decisive role for the I-CDW.\@

In the triclinic phase we have four different Ni-As bond distances of 2.320, 2.346, 2.360, and 2.369 {\AA} for Ni122 and of 2.313, 2.332, 2.356, and 2.384 {\AA} for BaNi$_2$(As,P)$_2$,\@ i.e.,\@ a more or less comparable splitting of the distances for the two compounds. The four significantly different Ni-Ni distances of the triclinic phase are shown in Table I and have already been discussed there. Comparing the intensity of the C-CDW peaks of Ni122 in Fig.\@ \ref{figS3}(a) and (b) and of the C-CDW peaks of BaNi$_2$(As,P)$_2$ in Fig.\@ 3(a) and (b),\@ the intensity of the C-CDW peaks (normalized to their corresponding fundamental reflections) is significantly increased with an increasing degree of distortion of the zig-zag chains: The more 'dimerized' character the Ni-Ni bonds have, the higher the intensity of the C-CDW.\@

\subsection{NEXAFS of BaNi$_2$As$_2$}

\begin{figure}[t]
\hspace{-3.0mm}
\includegraphics[width=0.4925\textwidth]{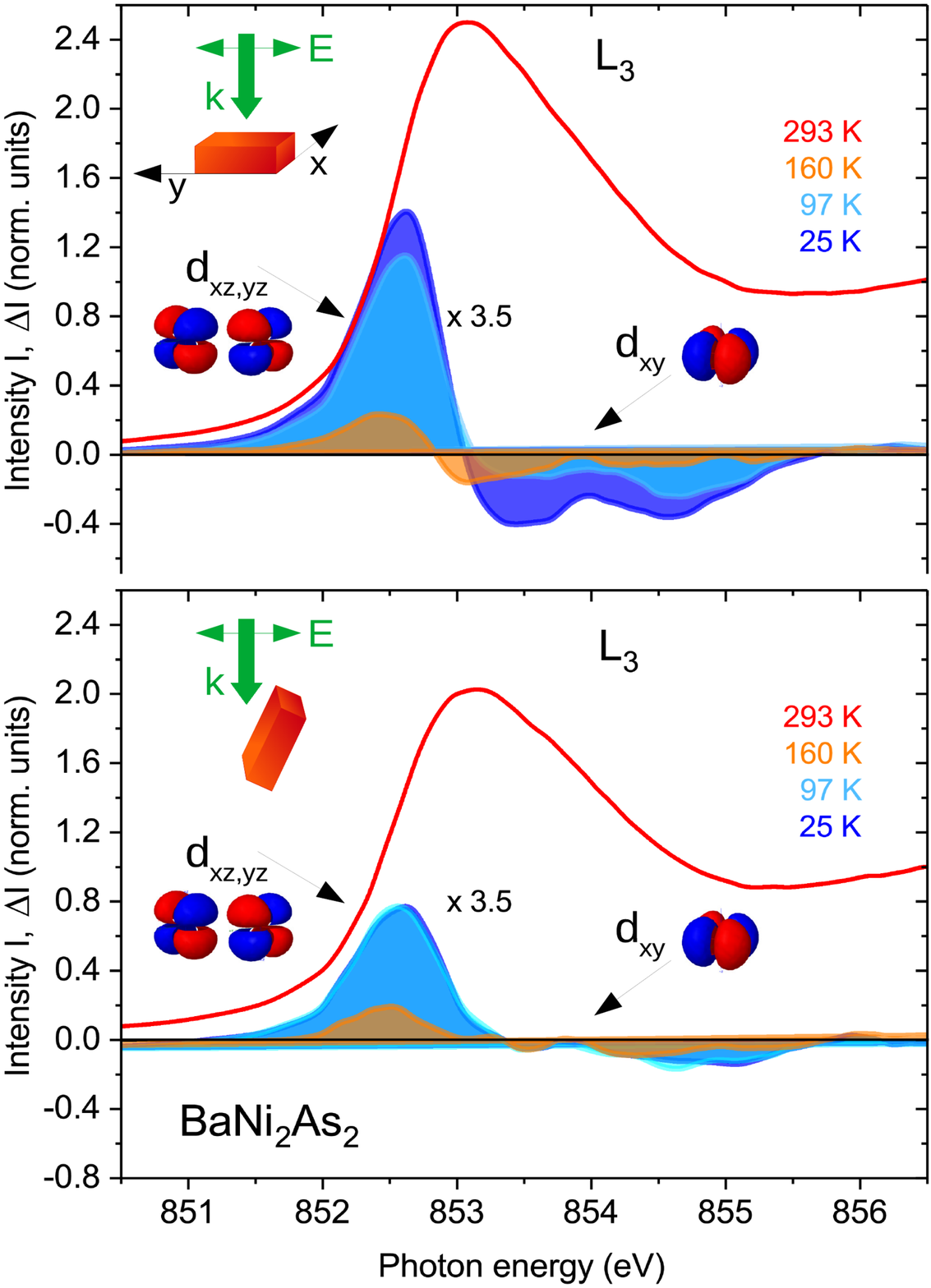}
\caption{\label{fig5}(Color online) Normal and grazing incidence NEXAFS of BaNi$_2$As$_2$.\@ Shown is the spectrum taken at 293 K and the difference between the spectra taken at 293 K and at the respective temperatures given in the graph. Please note that the difference is multiplied by a factor of 3.5.\@ In order to see the original spectra of this data set depicted, please refer to Fig.\@ \ref{figS15}.\@ A significant charge transfer from orbitals with in- and out-of-plane character, i.e., $d_{xz,yz}$ orbitals, to orbitals with exclusively in-plane character, i.e., to $d_{xy}$ orbitals is found below $T_{\rm S,2} \approx 137$ K. First indications for this transfer are, however, observed already far above $T_{\rm S,1}$.\@}
\end{figure}
Finally, we want to investigate if and how the charge ordering effects and the structural properties found for Ni122 are connected to the element-specific  orbital physics. As outlined above, NEXAFS is an ideal tool for determining the orbital-dependent unoccupied density of states. Using linearly polarized light, normal incidence probes predominantly orbitals with complete ($d_{xy}$, $d_{x^2 - y^2}$) or partial ($d_{xz,yz}$) in-plane character, whereas grazing incidence probes orbitals with predominantly out-of-plane contributions ($d_{xz}$, $d_{yz}$, $d_{3z^2 - r^2}$) and the observed spectral weight can be interpreted as the unoccupied density of states of the five $3d$ orbitals with the relevant directional characteristics. Fig.\@ \ref{fig5} depicts the temperature-dependent changes of the normal and grazing incidence NEXAFS data measured at the Ni $L_3$ edge of Ni122.\@ 
Taking the temperature-dependent changes of the difference spectra, $\Delta I = I$(293 K)$-I$($T$),\@ for normal and grazing incidence into account this demonstrates that for temperatures below $T_{\rm S,2} \approx 137$ K,\@ charge carriers are transferred from orbitals with simultaneous in- and out-of-plane character, i.e., $d_{xz,yz}$ orbitals (energy region  $\approx 852$ eV $\leq$ E $\leq 853$ eV), to orbitals with exclusively in-plane character, i.e., to $d_{xy}$ orbitals (energy region  $\approx 853$ eV $\leq$ E $\leq 855$ eV).\@ [Following Ref.\@ \cite{Merz2016} it is clear that the $d_{x^2 - y^2}$ and $d_{3z^2 - r^2}$ orbitals with $e$ symmetry are located at lower energy, are essentially filled, and do not contribute significantly to the spectral weight.] As can be seen, this effect is more pronounced for Ni122 (see Fig.\@ \ref{fig5}) than for BaNi$_2$(As,P)$_2$ (Fig.\@ 4).\@ By the same token, this transfer of charge carriers shows that, in addition to $d_{xz,yz}$ states, $d_{xy}$ orbitals are indispensable for establishing the Ni-Ni zig-zag chains and the C-CDW found in our XRD data on Ni122.\@ Yet the involvement of the $d_{xy}$ orbitals is reduced when going to BaNi$_2$(As,P)$_2$ (see Fig.\@ \ref{fig5p}).\@ Interestingly, a non-negligible amount of charge carriers is transferred between $d_{xz,yz}$ and $d_{xy}$ orbitals already at 160 K,\@ i.\@ e.\@,\@ far above the orthorhombic phase transition at $T_{\rm S,1} = 142$ K. This finding strongly points to possible charge/orbital fluctuations already in this $T$ range and is consistent with the diffuse scattering found in Fig.\@ \ref{figS2} which might be attributed to nematic effects as well. 
\begin{figure}[t]
\hspace{-7mm}
\includegraphics[width=0.5\textwidth]{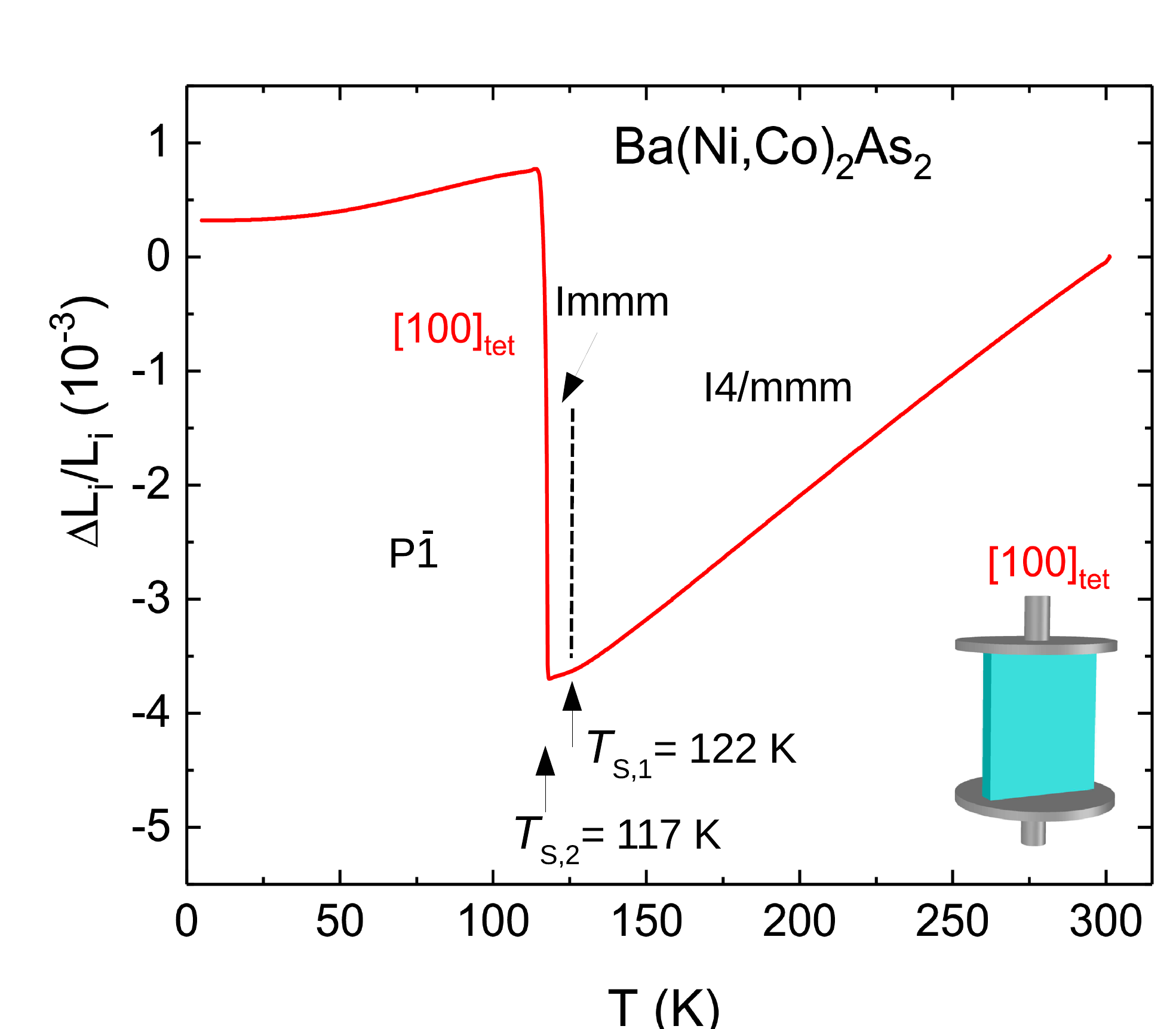}
\caption{\label{figS6} (Color online) Relative length change, $\Delta L/L$, versus temperature of the [100]$_{\rm tet}$ direction in the high-$T$ tetragonal notation for Ba(Ni,Co)$_2$As$_2$ obtained using high-resolution capacitance dilatometry. The whole measured temperature range is plotted in the figure.}
\end{figure}

We note that NEXAFS data of Ni122 illustrated in Fig.\@ \ref{fig5} show a stronger intensity difference between normal and grazing incidence than the corresponding spectra measured on BaNi$_2$(As,P)$_2$ (Fig.\@ \ref{fig5p}).\@ All of the data presented here are measured in fluorescence yield (FY). A big advantage of FY is that it is a bulk-sensitive method. On the other hand, this method sometimes suffers from significant self-absorption and saturation effects which can strongly distort the spectrum of white lines, especially when the saturation effects become more important for high-intensity grazing incidence measurements. This might be the case for our Ni122 grazing spectra. For BaNi$_2$(As,P)$_2$ the significant P background absorption due to the lower-lying P edges improves the situation, thereby leading to reduced differences between normal and grazing incidence. For our discussion, however, only the qualitative effect of redistribution of spectral weight between different orbital types is essential and the absolute magnitude of the spectral weight is less important. A possible additional explanation for a stronger intensity difference between normal and grazing incidence will be given below.

\section{\label{B} Measurements on Ba(Ni,Co)$_2$As$_2$}

\subsection{Thermal expansion of Ba(Ni,Co)$_2$As$_2$}

Fig.\@ \ref{figS6} presents the high-resolution thermal expansion, $\Delta L/L$,\@ of Ba(Ni,Co)$_2$As$_2$ measured along the [100]$_{\rm tet}$  direction in the high-$T$ tetragonal notation.\@ The orientation of the sample is sketched in the inset of Fig.\@ \ref{figS6}.\@ A pronounced first-order transition is observed below $T_{\rm S,2} \approx 117$ K.\@  Unfortunately, the sample could not be measured along the [110]$_{\rm tet}$ direction. Nevertheless, $T_{\rm S,1}$ was determined from the first derivative of $\Delta L/L$ to 122 K (not shown) which corroborates that the orthorhombic distortion exists for Ba(Ni,Co)$_2$As$_2$ as well.

\subsection{X-ray diffraction on Ba(Ni,Co)$_2$As$_2$}

\subsubsection{Temperature dependence of the (${\bar 3}$ ${\bar 3}$ ${\bar 6}$)$_{\rm tet}$ fundamental reflection}

\begin{figure}[b]
%\hspace{-2.5mm}
\includegraphics[width=0.475\textwidth]{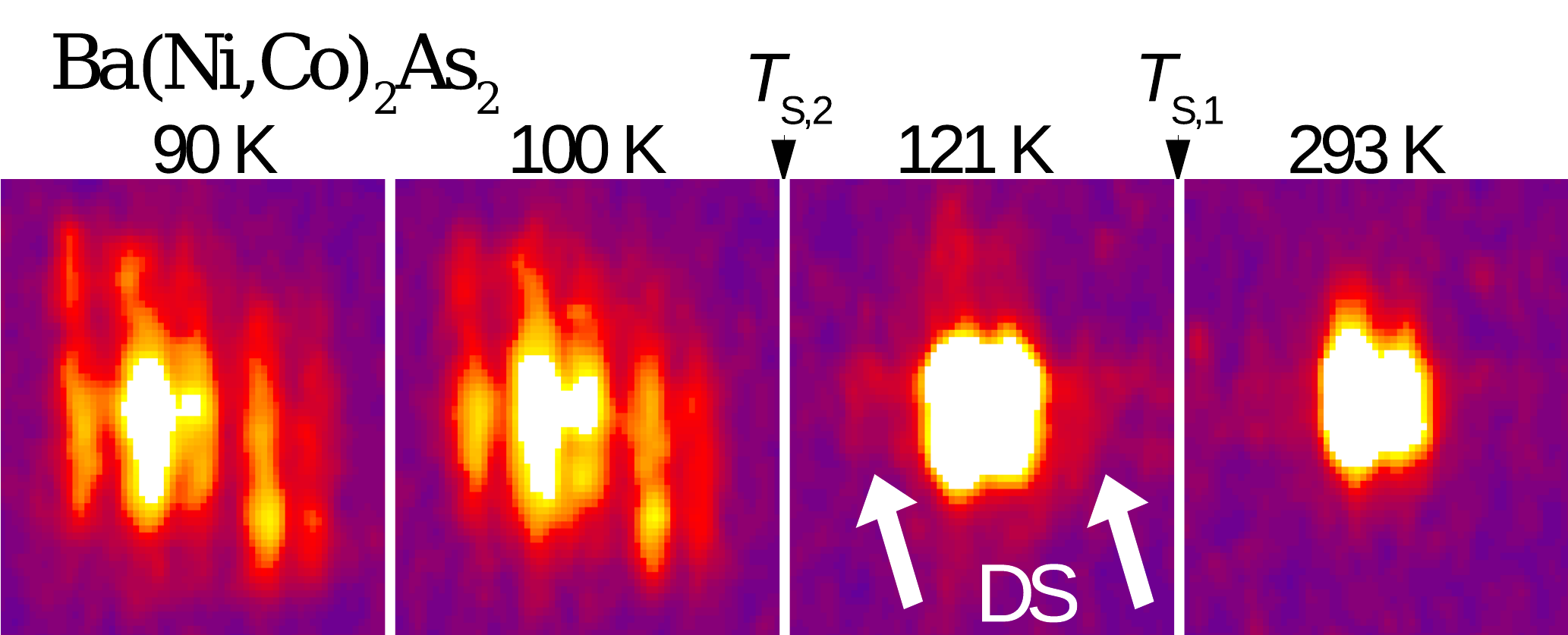}
\caption{\label{figS7} (Color online) Splitting of the (${\bar 3}$ ${\bar 3}$ ${\bar 6}$)$_{\rm tet}$ reflection at the structural phase transition of Ba(Ni,Co)$_2$As$_2$.\@ The structural transition to triclinic is clearly resolved by the peak splitting below $T_{\rm S,2}$,\@ the orthorhombic/nematic transition below $T_{\rm S,1}$ is indicated by an additional peak broadening. Despite an increased mosaic spread the increasing diffuse scattering upon cooling is still visible.} 
\end{figure}

The orthorhombic symmetry breaking derived from the TE data is supported by the additional peak broadening of the (${\bar 3}$ ${\bar 3}$ ${\bar 6}$)$_{\rm tet}$ reflection below $T_{\rm S,1}$  illustrated in Fig.\@ \ref{figS7}.\@ Finally, below $T_{\rm S,2} \approx 117$ K a triclinic peak splitting is observed which is stronger than the one of BaNi$_2$(As,P)$_2$ but comparable to the one of Ni122.\@ The shape and the width of the reflections demonstrates that our Ba(Ni,Co)$_2$As$_2$ sample has a slightly larger mosaic spread compared to the Ni122 and Ba(Ni,Co)$_2$As$_2$ samples which makes it more difficult to detect the weak but still visible diffuse scattering.

\subsubsection{Temperature dependence of the two types of charge density waves}

\begin{figure}[t]
\hspace{-5mm}
\includegraphics[width=0.5\textwidth]{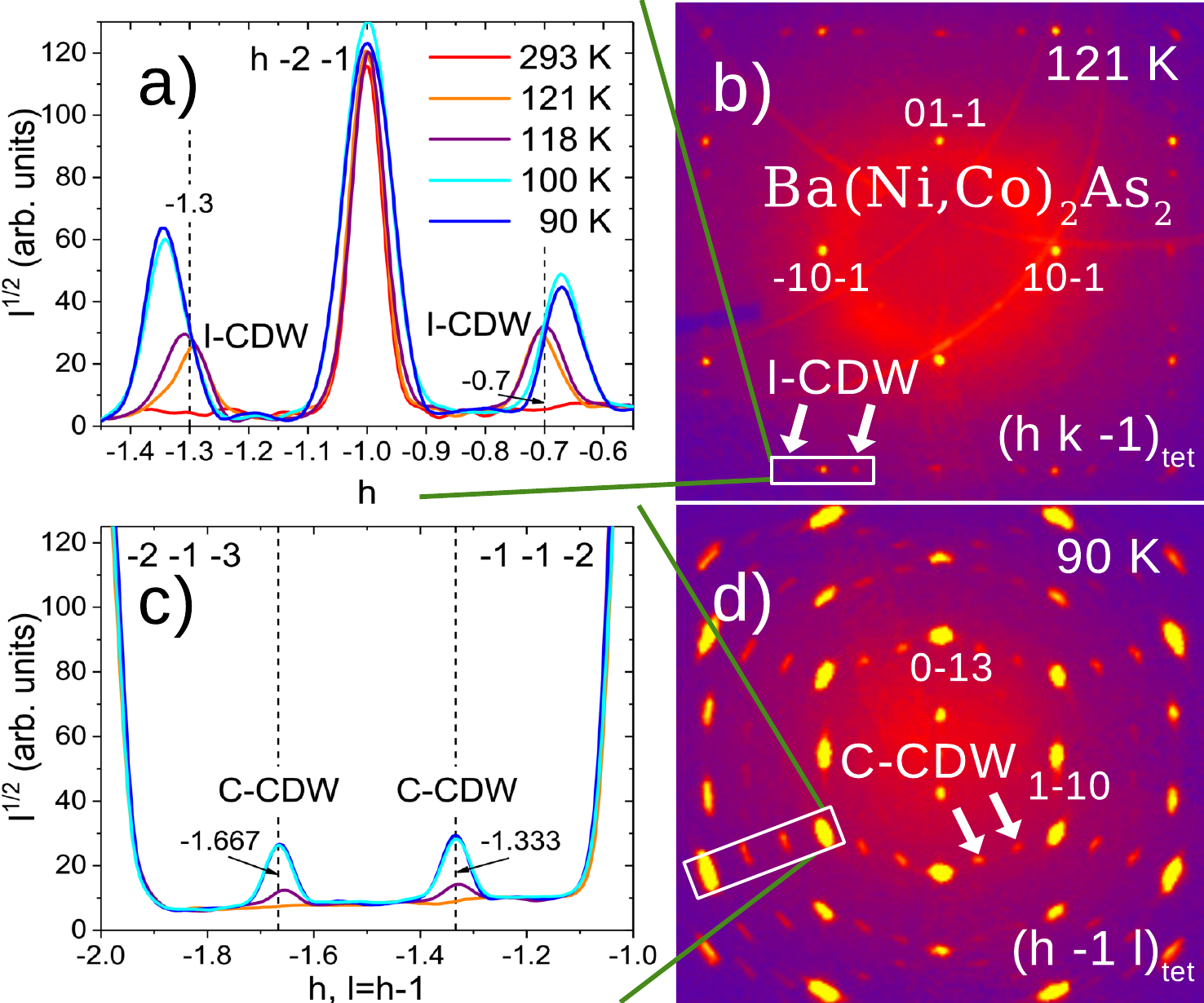}\\
\caption{\label{figS8}(Color online) (a) Temperature dependence of an ($h$ $k$ ${\bar 1}$)$_{\rm tet}$ plane of the reciprocal lattice for Ba(Ni,Co)$_2$As$_2$.\@ Below $T_{\rm S,1} \approx 122$ K incommensurate ($h \pm 0.28$ $k$ $l$)$_{\rm tet}$ and ($h$ $k \pm 0.28$ $l$)$_{\rm tet}$ superstructure reflections appear around the fundamental reflections and become blurry and split below $T_{\rm S,2} \approx 117$ K. (b) Temperature dependence of an ($h$ ${\bar 1}$ $l$)$_{\rm tet}$ plane of the reciprocal lattice for Ba(Ni,Co)$_2$As$_2$.\@ Below $T_{\rm S,2} \approx 117$ K commensurate ($h  \pm \frac{1}{3}$ $k$ $l \mp \frac{1}{3}$)$_{\rm tet}$ superstructure reflections appear around the fundamental reflections. In (a) and (b) arrows mark the positions of representative superstructure reflections.}
\end{figure}

Figs.\@ \ref{figS8}(a) and (b) present the temperature dependence of an ($h$ $k$ ${\bar 1}$)$_{\rm tet}$ plane of the reciprocal lattice for Ba(Ni,Co)$_2$As$_2$.\@ In the orthorhombic phase below $T_{\rm S,1} \approx 122$ K, incommensurate ($h \pm 0.28$ $k$ $l$)$_{\rm tet}$ superstructure reflections of the I-CDW, and due to the twinning, ($h$ $k \pm 0.28$ $l$)$_{\rm tet}$ ones appear around the fundamental reflections. In contrast to BaNi$_2$(As,P)$_2$,\@ however, the incommensurate ($h \pm 0.28$ $k$ $l$)$_{\rm tet}$ superstructure reflections are slightly weaker for Ba(Ni,Co)$_2$As$_2$ but still stronger than the ones of Ni122 in Fig.\@ \ref{figS3}(a).\@ In the triclinic phase below $T_{\rm S,2} \approx 117$ K the superstructure reflections smear out and move from $h = 0.28$ towards $\frac{1}{3}$ \cite{footnote16}.\@ Upon further cooling one of the CDW reflections around the fundamental reflections intensifies at the expense of the other which is attributed to the twinning of our samples.

To complete the picture of the charge ordering for Ba(Ni,Co)$_2$As$_2$,\@ the temperature dependence of an ($h$ ${\bar 1}$ $l$)$_{\rm tet}$ plane is illustrated in Figs.\@ \ref{figS8}(c) and (d).\@ Weak I-CDW reflections are observed below $T_{\rm S,1}$ for the ($h$ ${\bar 2}$ $l$)$_{\rm tet}$ plane (not shown).\@  Due to the larger mosaic spread observed for the investigated sample it is unclear if the I-CDW peaks disappear below $T_{\rm S,2}$ and we cannot completely rule out the phase coexistence reported in Ref.\@ \cite{Lee2019}.\@ At least at 118 K, i.\@ e.,\@ close to $T_{\rm S,2} = 117$ K,\@ Figs.\@ \ref{figS8}(a) and (c) point to the coexistence of both CDWs in a small $T$ range.\@ In the case of Ni122 and BaNi$_2$(As,P)$_2$,\@ however, we find absolutely no indication for phase coexistence in a wide temperature range. Yet the commensurate ($h \pm \frac{1}{3}$ $k$ $l \mp \frac{1}{3}$)$_{\rm tet}$ superstructure reflections appear around the fundamental reflections below $T_{\rm S,2}$.\@ Consistent with Ref.\@ \cite{Lee2019} we attribute the ($h \pm \frac{1}{3}$ $k$ $l \mp \frac{1}{3}$)$_{\rm tet}$ reflections to a commensurate charge density wave (C-CDW). By comparing the temperature dependence of the two types of CDWs between the three investigated systems it seems to be evident that the intensity of the I-CDW scales with the orthorhombic distortion whereas the intensity of the C-CDW seems to scale with an increasing  'dimerized' character of the Ni-Ni bonds (see also discussion in the following section and in Appendix \ref{AppA2c}).\@

\subsubsection{XRD: Structure determination and refinement}

Structural details of Ba(Ni,Co)$_2$As$_2$ resulting from refinements of diffraction data measured at temperatures characteristic for the tetragonal, orthorhombic/nematic, and triclinic phase, respectively, are listed in Table \ref{tab:table3}.\@ Since a complete crystallographic discussion is beyond the scope of an Appendix we will take a closer look only on the changes of the bond distances at the phase transitions.\@ In the tetragonal phase we have one Ni-Ni and one Ni-As bond distance of 2.924 and 2.359 {\AA},\@ respectively. In the orthorhombic phase we still have one Ni-Ni bond length of 2.917 {\AA};\@ the Ni-As bond length, however, splits up to 2.340 and 2.371 {\AA}.\@ Thus, consistent with our TE data we find a significant orthorhombic distortion for Ba(Ni,Co)$_2$As$_2$.\@ As discussed in the main paper, this splitting is responsible for lifting the degeneracy of the $d_{xz}$ and $d_{yz}$ orbitals and, thus, for the staggered ordering of the $d_{xz}$ orbitals along the crystallographic $a$ axis (see also Fig.\@ \ref{fig6}).\@ As mentioned already above, the degree of orthorhombic distortion, reflected by the magnitude of the Ni-As bond length splitting, is directly correlated with the intensity of the I-CDW superstructure reflections.\@ This finding suggests that the staggered ordering of the $d_{xz}$ orbitals plays a decisive role for the I-CDW.\@ 
In the triclinic phase we have four different Ni-As bond distances of 2.313, 2.332, 2.356, and 2.384 {\AA}.\@ The four significantly different Ni-Ni distances of the triclinic phase are shown in Table I and the consequences for zig-zag chains resulting from these differences have already been discussed there.

\begin{table}[t]
\caption{\label{tab:table3} Crystallographic data for Ba(Ni,Co)$_2$As$_2$ at 295 K,\@ 121 K,\@ and 90 K determined from single-crystal x-ray diffraction. The Co content was determined to 5 \%.\@ The high-temperature structure was refined in the tetragonal space group (SG) $I4/mmm$,\@ the nematic phase was refined in the orthorhombic SG $Immm$,\@ and the low-temperature structure was refined in the triclinic SG $P$${\bar 1}$.\@ $U_{\rm iso}$ denotes the isotropic atomic displacement parameters (ADP). The ADPs were refined anisotropically but due to space limitations only the $U_{\rm iso}$ are listed in the table. The Wyckoff positions (Wyck.) are given for their respective space groups. TW represents the degree of twinning in the corresponding phases. Statistical errors are shown.}
\begin{ruledtabular}
		\begin{tabular}[t]{llccc}
            &                           &       295 K          &          121 K      &       90 K \\ \hline
			& SG                        &     $I4/mmm$         &      $Immm$         & $P$${\bar 1}$    \\
			&  $a$ ({\AA})              &      4.1347(9)       &        4.1237(10)    &       4.1503(14) \\
			&  $b$ ({\AA})              &      4.1347(9)       &        4.1241(10)    &       4.1518(15) \\
			&  $c$ ({\AA})              &      11.6526(39)     &       11.6569(39)   &       6.4548(24) \\
			&  $\alpha$ (\textdegree)             &       90             &          90         &      108.502(27)   \\
			&  $\beta$ (\textdegree)              &       90             &          90         &      108.690(27)   \\
			&  $\gamma$ (\textdegree)             &       90             &          90         &       90.147(27) \\
			&  V ({\AA}$^3$)            &      199.2           &         198.2       &       99.0 \\ 
    Ba      & Wyck.                     &      $2a$            &         $2a$        &       $1a$ \\
			& $x$                       &        0             &          0          &         0 \\
			& $y$                       &        0             &          0          &         0 \\
			& $z$                       &        0             &          0          &         0 \\
			& $U_{\rm iso}$ ({\AA}$^2$) &      0.02727(32)     &      0.02179(99)   &     0.02201(61) \\ 
    Ni/Co   & Wyck.                     &       $4d$           &         $4j$        &     $2i$  \\
            & $x$                       &        $\frac{1}{2}$         &        $\frac{1}{2}$        &      0.76680(108) \\
			& $y$                       &          0           &           0         &     0.25401(143) \\
			& $z$                       &        $\frac{1}{4}$         &        0.24732(56)     &      0.49777(119) \\
			& $U_{\rm iso}$ ({\AA}$^2$) &  0.03333(44)         &      0.02746(116)    &      0.05215(168) \\
    As      & Wyck.                     &      $4e$            &         $4i$        &       $2i$ \\
            & $x$                       &        0             &          0          &      0.62824(54)\\
			& $y$                       &        0             &          0          &      0.65844(82) \\
			& $z$                       &    0.34757(13)        &      0.34770(12)    &      0.31153(34) \\
			& $U_{\rm iso}$ ({\AA}$^2$) &    0.03111(38)       &      0.02611(125)       &      0.03233(91) \\
			& TW (\%)                   &      $-$                &         56/44     &      34/25/23/18 \\
			& GOF                       &       2.93           &         2.75        &       3.01 \\
			& $wR_2$ (\%)               &       6.41           &         6.77        &       8.88  \\
			& $R_1$ (\%)                &       2.93           &         2.97        &       4.48  \\
		\end{tabular}
\end{ruledtabular}
\end{table}
\begin{table}[t]
\caption{\label{tab:table4} Crystallographic data for BaFe$_2$As$_2$ at 295 K,\@ 149 K,\@ and 90 K determined from single-crystal x-ray diffraction. The high-temperature structure was refined in the tetragonal space group (SG) $I4/mmm$ and the nematic phase was refined in the orthorhombic SG $Fmmm$.\@ For SG $Fmmm$,\@ $a$ and $b$ were estimated from the splitting of all symmetry equivalents of the (3 3 0)$_{\rm tet}$ reflection. $U_{\rm iso}$ denotes the isotropic atomic displacement parameters (ADP). The ADPs were refined anisotropically but due to space limitations only the $U_{\rm iso}$ are listed in the table. The Wyckoff positions (Wyck.) are given for their respective space groups. TW represents the degree of twinning in the corresponding phases. Statistical errors are shown.}
\begin{ruledtabular}
		\begin{tabular}[t]{llccc}
            &                           &       295 K          &          149 K      &       90 K \\ \hline
			& SG                        &     $I4/mmm$         &       $I4/mmm$      &      $Fmmm$    \\
			&  $a$ ({\AA})              &      3.9609(5)       &        3.9571(4)    &       5.6109(11) \\
			&  $b$ ({\AA})              &      3.9609(5)       &        3.9571(4)    &       5.5651(12) \\
			&  $c$ ({\AA})              &      13.0164(23)     &       12.9498(19)   &       12.9505(24) \\
			&  $\alpha$ (\textdegree)             &       90             &          90         &          90   \\
			&  $\beta$ (\textdegree)              &       90             &          90         &          90   \\
			&  $\gamma$ (\textdegree)             &       90             &          90         &          90    \\
			&  V ({\AA}$^3$)            &      204.2           &         202.8       &       405.5 \\ %%\hline
    Ba      & Wyck.                     &      $2a$            &         $2a$        &       $4a$ \\
			& $x$                       &        0             &          0          &         0 \\
			& $y$                       &        0             &          0          &         0 \\
			& $z$                       &        0             &          0          &         0 \\
			& $U_{\rm iso}$ ({\AA}$^2$) &      0.01336(13)     &      0.00847(14)   &     0.00894(86) \\ %%\hline
    Fe      & Wyck.                     &       $4d$           &         $4d$        &      $8f$  \\
            & $x$                       &        $\frac{1}{2}$         &        $\frac{1}{2}$        &      $\frac{1}{4}$ \\
			& $y$                       &          0           &           0         &      $\frac{1}{4}$ \\
			& $z$                       &        $\frac{1}{4}$         &        $\frac{1}{4}$        &      $\frac{1}{4}$ \\
			& $U_{\rm iso}$ ({\AA}$^2$) &  0.010883(17)         &      0.00755(19)    &      0.00865(115) \\
    As      & Wyck.                     &      $4e$            &         $4e$        &       $8i$ \\
            & $x$                       &        0             &          0          &         0    \\
			& $y$                       &        0             &          0          &         0    \\
			& $z$                       &    0.35411(5)        &      0.35383(6)    &      0.35386(7) \\
			& $U_{\rm iso}$ ({\AA}$^2$) &    0.01094(15)       &      0.00745(17)       &    0.00821(90) \\
			& TW (\%)                   &      $-$                &         $-$     &      52/48 \\
			& GOF                       &       1.73           &         1.88        &       1.92 \\
			& $wR_2$ (\%)               &       3.72           &         4.07        &       4.66  \\
			& $R_1$ (\%)                &       1.31           &         1.50        &       1.91  \\
		\end{tabular}
\end{ruledtabular}
\end{table}

\subsection{NEXAFS of Ba(Ni,Co)$_2$As$_2$}

\begin{figure}[t]
\hspace{-3.0mm}
\includegraphics[width=0.4925\textwidth]{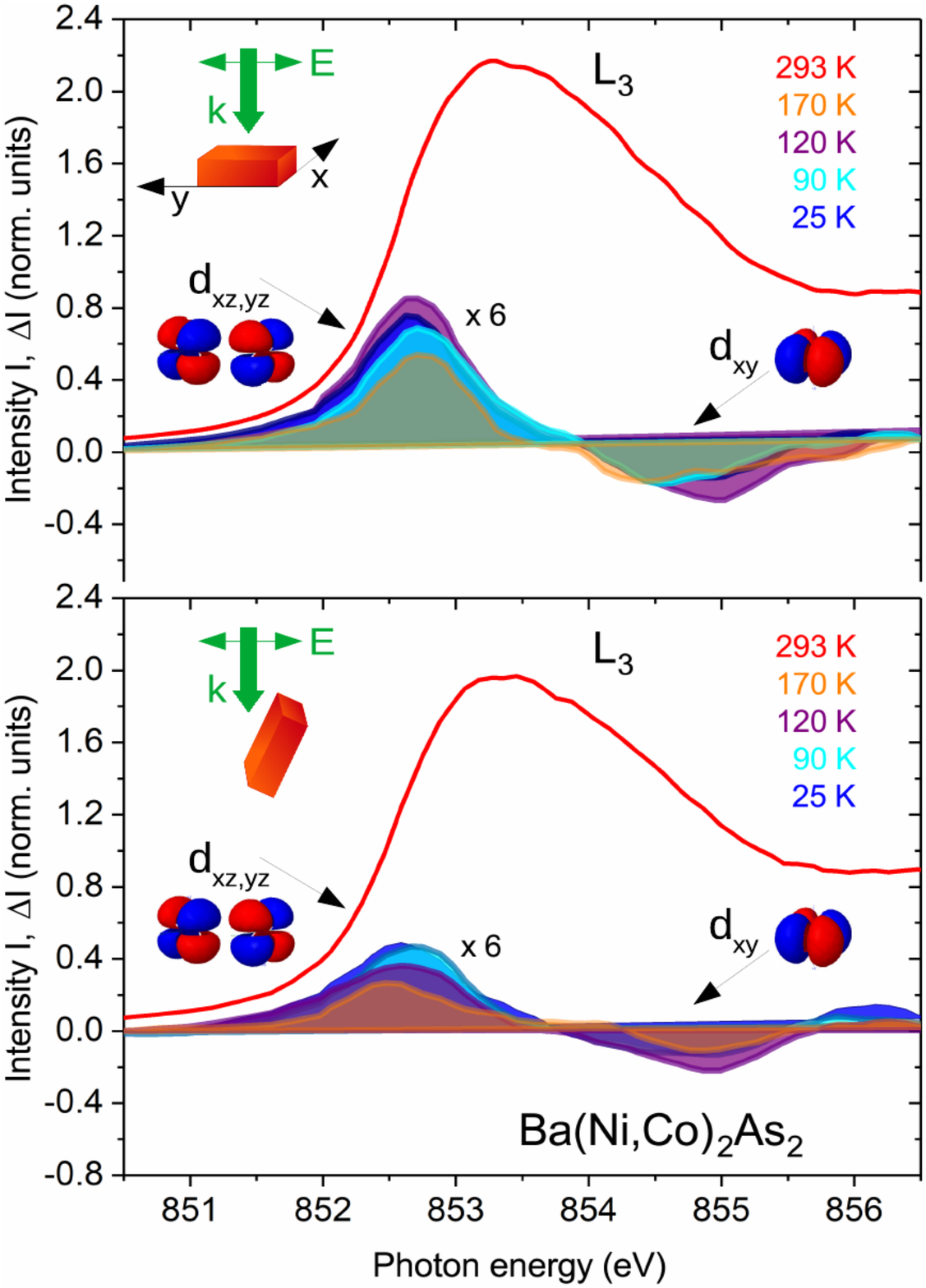}
\caption{\label{figS9}(Color online) Normal and grazing incidence NEXAFS of Ba(Ni,Co)$_2$As$_2$.\@ Shown is the spectrum taken at 293 K and the difference between the spectra taken at 293 K and at the respective temperatures given in the graph. Please note that the difference is multiplied by a factor of 6.\@ A significant charge transfer from orbitals with in- and out-of-plane character, i.e., $d_{xz,yz}$ orbitals, to orbitals with exclusively in-plane character, i.e., to $d_{xy}$ orbitals is found below $T_{\rm S,2} \approx 117$ K. First indications for this transfer are, however, observed already far above $T_{\rm S,1}$.\@}
\end{figure}

Fig.\@ \ref{figS9} depicts the temperature-dependent changes of the normal and grazing incidence NEXAFS data measured at the Ni $L_3$ edge of Ba(Ni,Co)$_2$As$_2$.\@ Judging by the statistical variations of the measuring points it is obvious that the relative error bars of the Ba(Ni,Co)$_2$As$_2$ spectra are somewhat larger than for the other two studied compounds --- however, without significantly affecting a qualitative interpretation of the data. Taking the temperature-dependent changes of the difference spectra, $\Delta I = I$(293 K)$-I$($T$),\@ for normal and grazing incidence into account this demonstrates that for temperatures below $T_{\rm S,2} \approx 117$ K,\@ charge carriers are transferred from orbitals with simultaneous in- and out-of-plane character, i.e., $d_{xz,yz}$ orbitals (energy region  $\approx 852$ eV $\leq$ E $\leq 853$ eV), to orbitals with exclusively in-plane character, i.e., to $d_{xy}$ orbitals (energy region  $\approx 854$ eV $\leq$ E $\leq 856$ eV).\@ [Following Ref.\@ \cite{Merz2016} it is clear that the $d_{x^2 - y^2}$ and $d_{3z^2 - r^2}$ orbitals with $e$ symmetry are located at lower energy, are essentially filled, and do not contribute significantly to the spectral weight.] Again, this transfer of charge carriers shows that, in addition to $d_{xz,yz}$ states, $d_{xy}$ orbitals are indispensable for establishing the (distorted) Ni-Ni zig-zag chains and the C-CDW found in our XRD data.\@ As can be seen, however, this effect is weaker for Ba(Ni,Co)$_2$As$_2$ than for Ni122 (see Fig.\@ \ref{fig5}) and BaNi$_2$(As,P)$_2$ (Fig.\@ \ref{fig5p}).\@ 

Similar to Ni122 a more pronounced intensity difference between normal and grazing incidence is observed than for BaNi$_2$(As,P)$_2$ (Fig.\@ \ref{fig5p}),\@ which might again be attributed to significant self-absorption and saturation effects (see also discussion for Ni122 above). On the other hand, the intensity difference between normal and grazing incidence could also be correlated with the (distorted) zig-zag chains: The more distorted the zig-zag chains become when going from Ni122 to Ba(Ni,Co)$_2$As$_2$ and finally to BaNi$_2$(As,P)$_2$ the smaller the contribution of the $d_{xy}$ orbitals will be and eventually the less charge carriers have to be transferred to these orbitals in order to facilitate the (distorted) zig-zag chains.

In addition to the transfer of spectral weight observed below $T_{\rm S,2}$ a comparable amount of charge carriers is, surprisingly, transferred between $d_{xz,yz}$ and $d_{xy}$ orbitals already at 120 K,\@ i.\@ e.\@,\@ above the triclinic phase transition at $T_{\rm S,2} = 117$ K. Since 120 K is only slightly above $T_{\rm S,2}$ we have considered that the Co content possibly differs somewhat between samples from the same batch and measured the sample at 170 K,\@ i.\@ e.\@,\@ far above the expected tetragonal-to-orthorhombic transition at $T_{\rm S,1} = 122$ K.\@ Yet also at this temperature a significant transfer of spectral weight between $d_{xz,yz}$ and $d_{xy}$ orbitals is found which again points to possible charge/orbital fluctuations already in this $T$ range. This finding is consistent with the diffuse scattering found in all of our XRD experiments which might be attributed to nematic effects.

\section{\label{AppC}Measurements on BaFe$_2$As$_2$}

Finally, we compare our measurements on BaNi$_2$(As,P)$_2$,\@ Ba(Ni,Co)$_2$As$_2$,\@ and Ni122 with the corresponding ones on BaFe$_2$As$_2$.

\subsection{Thermal expansion of BaFe$_2$As$_2$}

The high-resolution thermal expansion, $\Delta L/L$,\@ of BaFe$_2$As$_2$ shows a strong difference between the orthorhombic $a$ and $b$ axis below the tetragonal-to-orthorhombic phase transition which is well documented for instance in Refs.\@ \cite{Boehmer2015,Wang2016,Wang2018} and will not be shown here.

\subsection{X-ray diffraction on BaFe$_2$As$_2$}

\subsubsection{Temperature dependence of the (${\bar 3}$ ${\bar 3}$ ${\bar 6}$)$_{\rm tet}$ fundamental reflection}

\begin{figure}[b]
\includegraphics[width=0.475\textwidth]{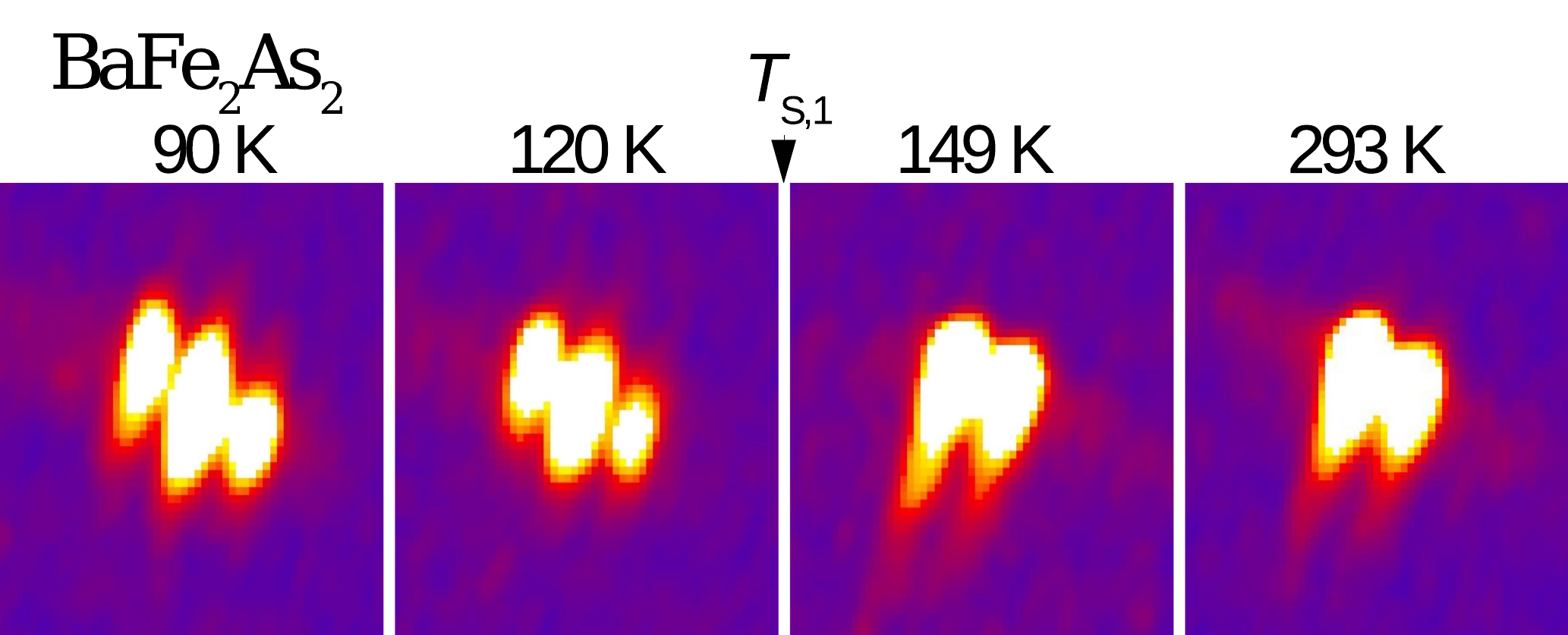}
\caption{\label{figS10} (Color online) Splitting of the (${\bar 3}$ ${\bar 3}$ ${\bar 6}$)$_{\rm tet}$ reflection at the structural phase transition of BaFe$_2$As$_2$.\@ The structural transition to orthorhombic is clearly resolved by the peak splitting below $T_{\rm S,1}$.\@} 
\end{figure}
\begin{figure*}[t]
\hspace{-5mm}
\includegraphics[width=0.935\textwidth]{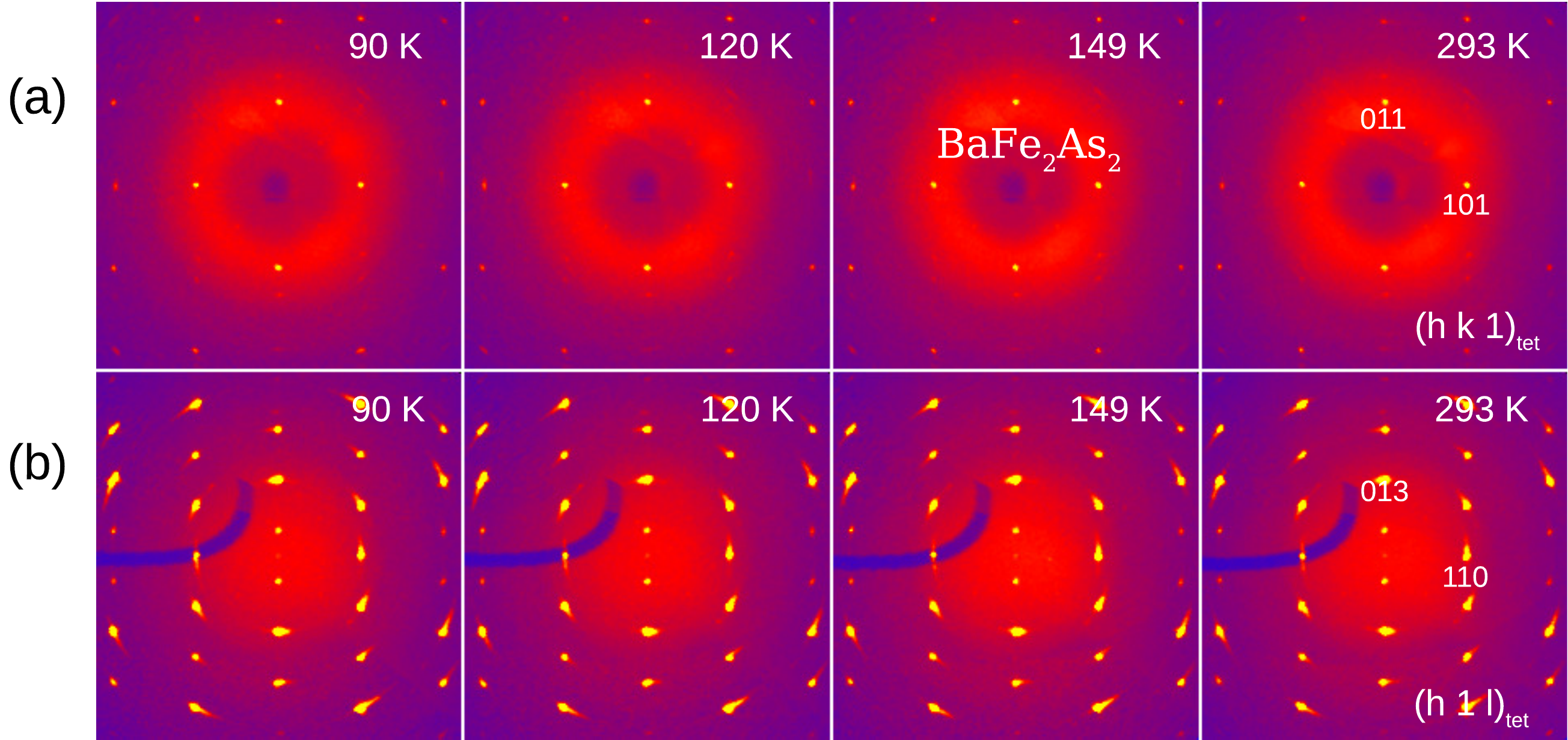}\\
\caption{\label{figS11}(Color online) (a) Temperature dependence of an ($h$ $k$ 1)$_{\rm tet}$ plane of the reciprocal lattice for BaFe$_2$As$_2$.\@ (b) Temperature dependence of an ($h$ 1 $l$)$_{\rm tet}$ plane of the reciprocal lattice for BaFe$_2$As$_2$.\@ No indication for I-CDW or C-CDW superstructure reflections are detected around the fundamental reflections.}
\end{figure*}
The orthorhombic symmetry breaking derived from the TE data is supported by the peak splitting of the (${\bar 3}$ ${\bar 3}$ ${\bar 6}$)$_{\rm tet}$ reflection below $T_{\rm S,1} \approx 135$ K illustrated in Fig.\@ \ref{figS10}.\@ The figure also illustrates how the thermal diffuse scattering (TDS) is reduced upon cooling when (nematic) orbital fluctuations are absent: For Fe122 the nematic transition is (without a final smoking-gun proof) very often believed to be induced by magnetic fluctuations which break the fourfold symmetry at $T_{\rm S}$ \cite{Fernandes2014}.\@

Since XRD is insensitive to \textit{magnetic} fluctuations they cannot be detected with x-rays in contrast to the \textit{charge} fluctuations found for BaNi$_2$(As,P)$_2$,\@ Ba(Ni,Co)$_2$As$_2$,\@ and Ni122 in our data.

\subsubsection{Temperature dependence of the two types of charge density waves}

For Fe122 so far only spin density waves but no charge density waves have been observed and this is reflected by the absence of any kind of superstructure reflections in our investigation of the reciprocal lattice. Just for completeness Fig.\@ \ref{figS11}(a) therefore presents the temperature dependence of an ($h$ $k$ $1$)$_{\rm tet}$ plane of the reciprocal lattice for Fe122 and Fig.\@ \ref{figS11}(b) the one of an ($h$ $1$ $l$)$_{\rm tet}$ plane.\@ In the whole temperature range no indications for incommensurate ($h \pm 0.28$ $k$ $l$)$_{\rm tet}$,\@ let alone, commensurate ($h \pm \frac{1}{3}$ $k$ $l \mp \frac{1}{3}$)$_{\rm tet}$ superstructure reflections around the fundamental reflections are found.

\subsubsection{XRD: Structure determination and refinement}

\begin{figure}[b]
\hspace{-3.8mm}
\includegraphics[width=0.495\textwidth]{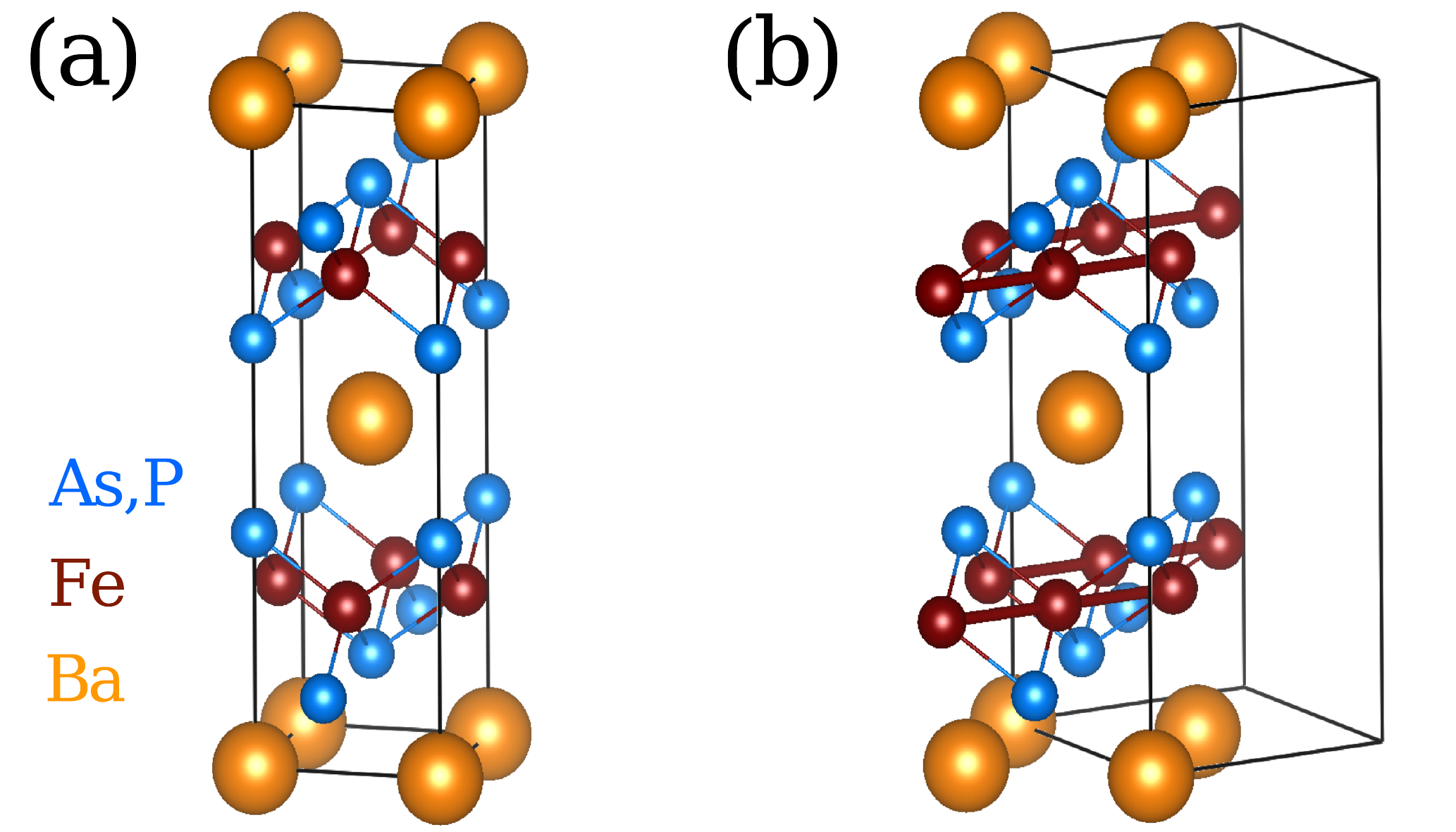}\\[-0.5mm]
\caption{\label{figS12}(Color online)  Spatial structure of Fe122 in the (a) high-$T$ tetragonal (SG $I4/mmm$) (b) low-$T$ orthorhombic/nematic (SG $Fmmm$) phase. The reduced Fe-Fe bond distances (represented by thicker lines) lead to an orbital ordering along the tetragonal [110]$_{\rm tet}$ direction induced by the lifting of the degeneracy of $d_{xz}$ and $d_{yz}$ orbitals. Please note that the unit cells of the tetragonal and the orthorhombic phase (thin black lines) are rotated against each other by 45\textdegree.\@}
\end{figure}
Structural details of Fe122 resulting from refinements of diffraction data measured at temperatures characteristic for the tetragonal and the orthorhombic/nematic phase are listed in Table \ref{tab:table4}.\@ Since a complete crystallographic discussion is beyond the scope of an Appendix we will take a closer look only on the changes of the bond distances at the phase transition.\@ In the tetragonal phase we have one Fe-Fe and one Fe-As bond distance of 2.801 and 2.400 {\AA},\@ respectively. In the orthorhombic phase we have two Fe-Fe bond lengths of 2.780 and 2.805 {\AA},\@ and one Fe-As bond length of 2.389 {\AA}.\@ The refinement of the data measured at 120 K is almost identical to the one of the 90 K data compiled in Table \ref{tab:table4},\@ which demonstrates that, apart from small changes in the lattice parameters, no significant structural changes occur in the orthorhombic phase upon further cooling. In Fig.\@ \ref{figS12} the corresponding unit cells of (a) the high-$T$ tetragonal and (b) the low-$T$ orthorhombic/nematic phase are depicted. In Fig.\@ \ref{figS12}(b) it is indicated, how the reduced Fe-Fe bond distances (represented by thicker lines) lead to the established orbital ordering along the tetragonal [110]$_{\rm tet}$ direction:\@ The difference in Fe-Fe bond lengths results in a lifting of the degeneracy of $d_{xz}$ and $d_{yz}$ orbitals and is most probably induced by spin-nematic fluctuations \cite{Fernandes2014}.

\subsection{NEXAFS of BaFe$_2$As$_2$}

\begin{figure}[t]
\hspace{-3.0mm}
\includegraphics[width=0.4925\textwidth]{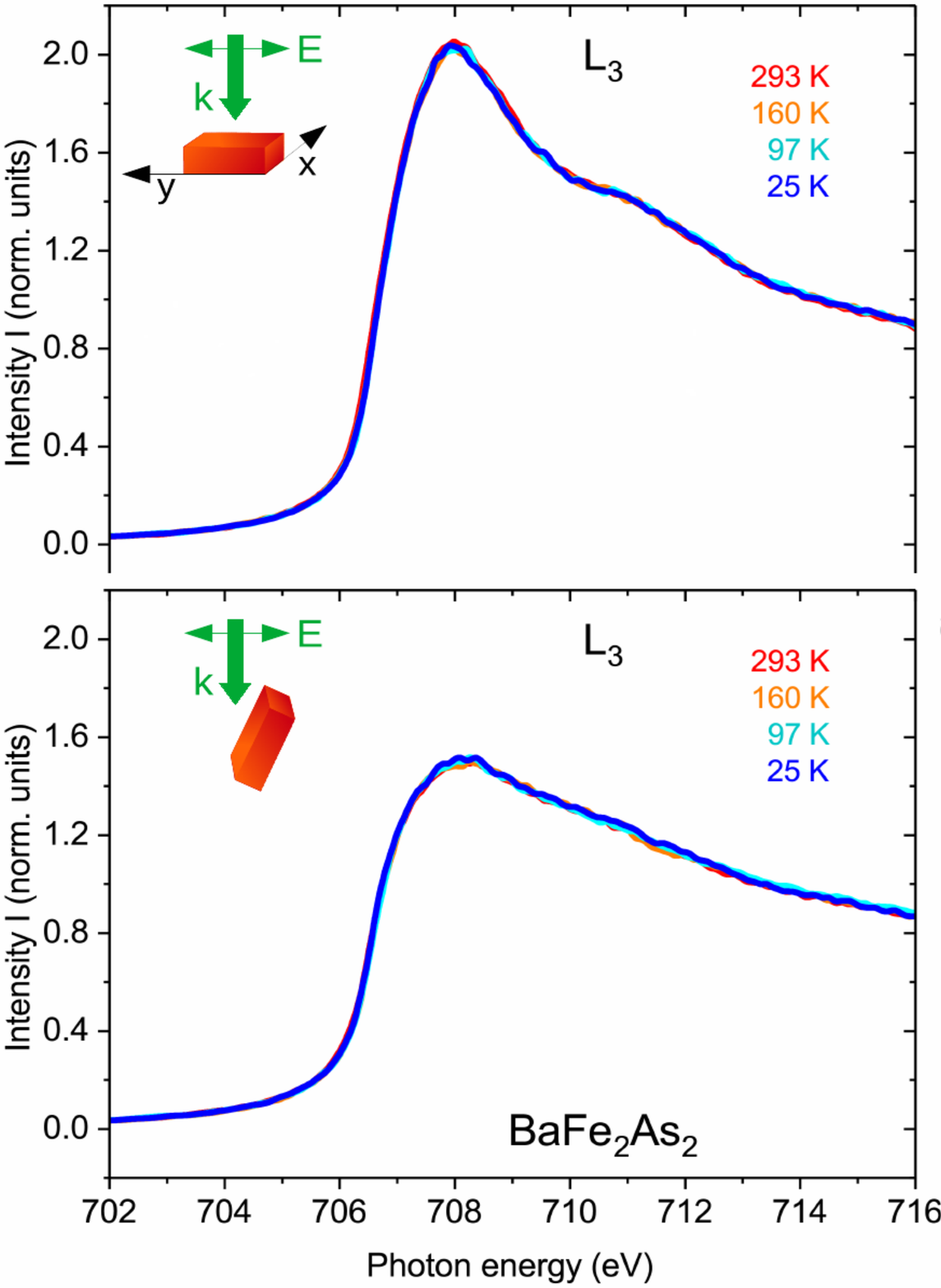}
\caption{\label{figS13}(Color online) Normal and grazing incidence NEXAFS of BaFe$_2$As$_2$.\@ Shown are the spectra taken at 293 K,\@ 160 K,\@ 97 K,\@ and 25 K.\@
In contrast to all other compounds investigated in this paper, all of the normal incidence and all of the grazing incidence spectra, respectively, are identical and fall on top of each other and, thus, no $T$-dependent charge transfer between out-of-plane and in-plane orbital types is observed for BaFe$_2$As$_2$.\@}
\end{figure}
\begin{figure}[t]
\vspace{1.0mm}
\includegraphics[width=0.4925\textwidth]{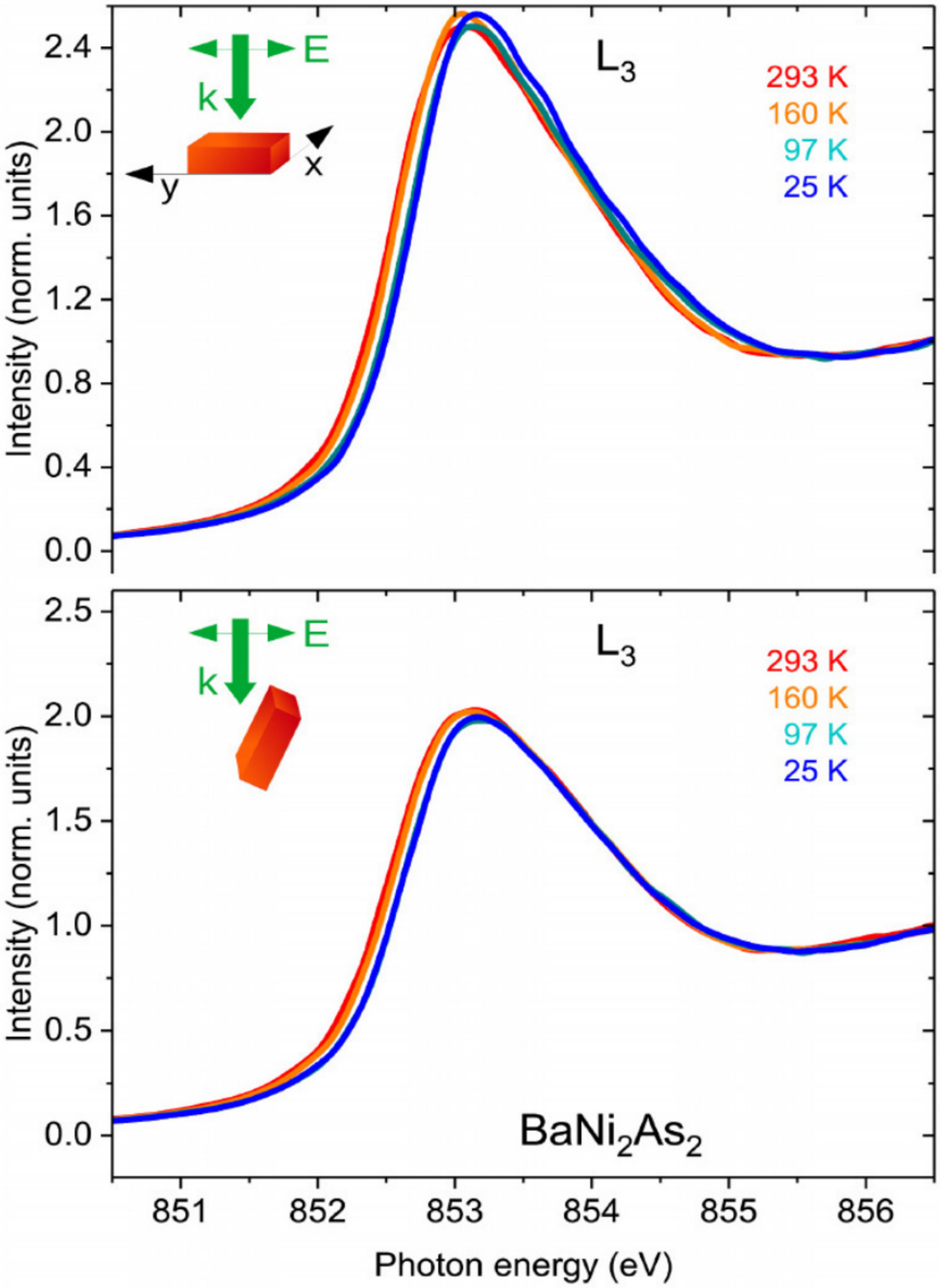}
\vspace{-5.0mm}
\caption{\label{figS15}(Color online) Normal and grazing incidence NEXAFS of BaNi$_2$As$_2$.\@ Shown are the spectra taken at 293 K,\@ 160 K,\@ 97 K,\@ and 25 K.\@ These are precisely the same data as presented in Fig.\@ \ref{fig5} --- just plotted as original spectra, not as difference spectra. In contrast to BaFe$_2$As$_2$, all of the normal incidence and all of the grazing incidence spectra, respectively, show a significant $T$-dependent change of the spectral weight and, thus, the $T$-dependent charge transfer between out-of-plane and in-plane orbital types discussed in detail above.\@}
\end{figure}
Fig.\@ \ref{figS13} depicts the temperature-dependent changes of the normal and grazing incidence NEXAFS data measured at the Fe $L_3$ edge of Fe122.\@ Spectra were measured in the teragonal phase at 293 and 160 K and in the orthorhombic phase at 97 and 25 K.\@ In contrast to the other compounds investigated in this paper, all of the normal incidence spectra and all of the grazing incidence spectra are identical and fall on top of each other and, thus, no $T$-dependent charge transfer between $d_{xz,yz}$ and $d_{xy}$ orbitals is observed for BaFe$_2$As$_2$.\@ In principle, small spectral changes could have been expected between the high- and low-$T$ data at the tetragonal-to-orthorhombic transition due the energetic splitting of the initially degenerate $d_{xz}$/$d_{yz}$ orbitals.\@ Most probably this effect is, however, washed out by (a) the strong twinning of our samples which averages out the spectral difference between the two twins and (b) the fact that NEXAFS measures the {\bf k}-integrated unoccupied density of states. More important, the spectra clearly show that the orbital type is not changed at the transition and a charge transfer between in- and out-of-plane orbitals as observed for Ni122,\@ BaNi$_2$(As,P)$_2$,\@ and  Ba(Ni,Co)$_2$As$_2$ can definitely be excluded. Fig.\@ \ref{figS13},\@ with its complete absence of any temperature-dependent redistribution of spectral weight for Fe122, is directly juxtaposed to Fig.\@ \ref{figS15} which clearly demonstrates, again, the presence of such redistributions for Ni122.\@

\section{\label{AppD}Spin-driven vs. orbital-driven nematicity}

\begin{figure}[b]
\hspace{-1.0mm}
\includegraphics[width=0.415\textwidth]{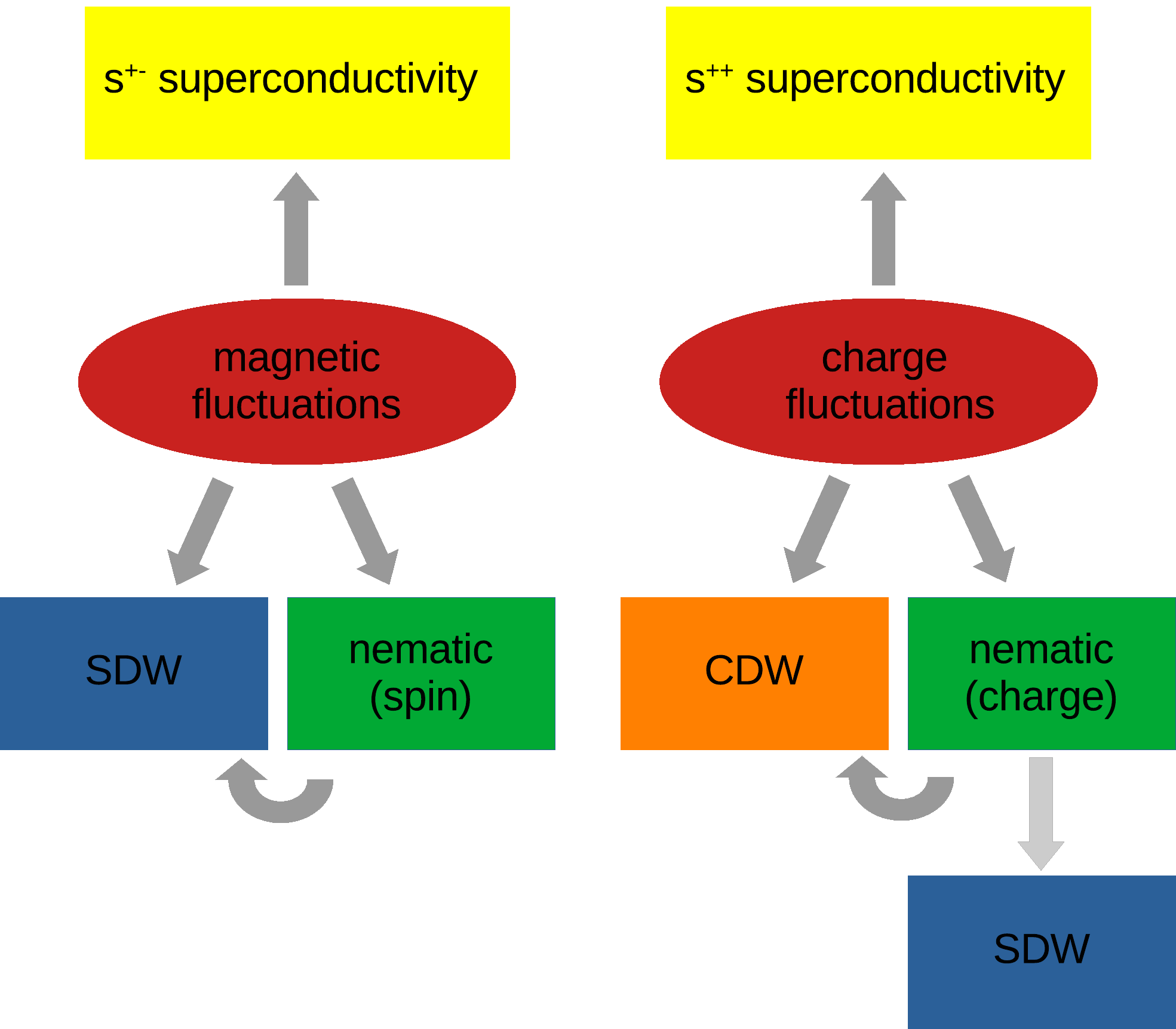}
\caption{\label{figS14}(Color online) Hierarchy of the electronic ordered states for magnetic and charge/orbital fluctuations adapted from Ref.\@ \cite{Fernandes2014}.\@}
\end{figure}

In this final chapter we want to discuss the differences between structural transitions for which the fourfold symmetry is broken (a) by magnetic fluctuations or (b) by charge/orbital fluctuations. For this purpose we have plotted the hierarchy of the electronic ordered states for both kinds of fluctuations in Fig.\@ \ref{figS14};\@ this hierarchy is adapted from Ref.\@ \cite{Fernandes2014}, Fig.\@ 5b.\@ The chapter largely follows the corresponding treatment in Refs.\@ \cite{Fernandes2014,Fernandes2012} but continuously juxtaposes it with the key structural and spectroscopic aspects of our present systems thus providing a first glimpse at how orbital-driven nematicity might come about in real-world systems such as Ni122,\@ BaNi$_2$(As,P)$_2$,\@ and Ba(Ni,Co)$_2$As$_2$.\@

(a) Magnetic-fluctuation-induced nematicity: The low-temperature SDW ground state of these systems  is characterized by magnetic stripes of parallel spins along either the crystallographic {\bf b} (ordering vector $Q_1 = (\pi, 0)$) or {\bf a} axis (ordering vector $Q_2 = (0, \pi)$) and an antiparallel ordering of the spins in the perpendicular in-plane direction \cite{Fernandes2014,Fernandes2012}.\@ For this ordered state the average spin at position $i$,\@ $\langle {\bf S}_i \rangle \neq$ 0 and for the spin correlations $\langle {\bf S}_i \cdot {\bf S}_{i+x} \rangle =  - \langle {\bf S}_i \cdot {\bf S}_{i+y} \rangle$ and the system has to break not only the $O(3)$ spin-rotational (time-reversal) symmetry, but it also has to choose between the two said degenerate ground states which corresponds to a $Z_2$ (Ising-like) symmetry breaking \cite{Fernandes2014,Fernandes2012}.\@ At RT, on the other hand, the structure of these systems has a paramagnetic  tetragonal symmetry with SG $I4/mmm$ for which $\langle {\bf S}_i \rangle =  0$ and $\langle {\bf S}_i \cdot {\bf S}_{i+x} \rangle =  \langle {\bf S}_i \cdot {\bf S}_{i+y} \rangle = 0$.\@ Upon cooling, the magnetic fluctuations increase and at $T_{\rm nem}$ they become strong enough to trigger a second-order nematic/structural transition to SG $Fmmm$.\@ In the temperature range $T_{\rm mag} \leq T \leq T_{\rm nem}$ of this nematic phase  $\langle {\bf S}_i \rangle = 0$ but the spins are correlated, i.e.,\@ $\langle {\bf S}_i \cdot {\bf S}_{i+x} \rangle =  - \langle {\bf S}_i \cdot {\bf S}_{i+y} \rangle$,\@ thereby breaking the fourfold symmetry.\@ As a consequence of the  $Z_2$ symmetry breaking the $d_{xz}$ and $d_{yz}$ orbitals are no longer degenerate and the spin-driven nematic order gives rise to orbital order below $T_{\rm nem}$.\@ Finally, the $O(3)$ spin-rotational symmetry is broken below $T_{\rm mag}$ as well, $\langle {\bf S}_i \rangle \neq  0$,\@ and the SDW ground state appears. A representative of these systems is Fe122,\@ especially in a certain substitution range when Fe is partially replaced with Co or Ni.\@ For these systems all of the phases predicted by the nematic model are experimentally found.

(b) Charge/orbital-fluctuation-induced nematicity: Even though the charge/orbital-fluctuation model is more complex it is, in its simplest form, completely analogous to the magnetic one, with the main difference that the interaction between electron and hole pockets is attractive while it is repulsive for magnetic fluctuations \cite{Fernandes2014}.\@ The ground state is a CDW with an ordering vector ${\bf Q}_X$ or ${\bf Q}_Y$ for which translation and $Z_2$ symmetry is broken. In the ground state the CDW order parameter $\langle W_i \rangle \neq 0$ and $\langle {W_X}^2 \rangle \neq \langle {W_Y}^2 \rangle$.\@ In the high-$T$ phase, on the other hand, the structure of these systems has a tetragonal symmetry with SG $I4/mmm$ for which $\langle W_i \rangle = 0$ and $\langle {W_X}^2 \rangle = \langle {W_Y}^2 \rangle = 0$.\@ Upon cooling, the charge fluctuations increase and at $T_{\rm nem}$ they become strong enough to trigger a second-order nematic/structural transition. In the temperature range $T_{\rm CDW} \leq T \leq T_{\rm nem}$ of the nematic phase  $\langle W_i \rangle = 0$ but $\langle {W_X}^2 \rangle \neq \langle {W_Y}^2 \rangle$ and the system spontaneously develops an orbital order and the fourfold symmetry is broken \cite{Fernandes2014}.\@ Finally the translation symmetry is broken below $T_{\rm CDW}$ as well, $\langle W_i \rangle \neq 0$,\@ and the CDW ground state appears.

Let us now compare the model described in (b) to what we observe experimentally for Ni122,\@ BaNi$_2$(As,P)$_2$,\@ and Ba(Ni,Co)$_2$As$_2$.\@ All of these systems have in the high-$T$ phase SG type $I4/mmm$.\@ Upon cooling we find in our NEXAFS data strong indications for charge fluctuations between out-of-plane and in-plane orbitals. Concomitantly, we find in our XRD data an increasing diffuse scattering which points to charge fluctuations as well.\@ At $T_{\rm S,1}$ which corresponds to $T_{\rm nem}$ in the theoretical model (see above) the fluctuations become strong enough and a structural transition to SG $Immm$ takes place for which the fourfold symmetry is broken. This $Z_2$ symmetry breaking together with the orbital ordering observed in our XRD studies is in line with a nematic transition. Interestingly an I-CDW accompanies this transition. The intensity of the I-CDW direcly scales with the orthorhombic distortion and the staggered ordering of the $d_{xz}$ orbitals is tied to the I-CDW.\@ Therefore, the finding of an I-CDW suggests an interpretation as a possible charge/orbital-driven electronic nematic phase.\@ The question of its origin, however, is not fully answered yet.\@ Finally, the transition to the triclinic SG $P$${\bar 1}$ and the C-CDW appear when cooled below $T_{\rm S,2}$ ($T_{\rm S,2}$ corresponds to $T_{\rm CDW}$ in the theoretical model described above).\@ While the finding of the C-CDW is expected for charge/orbital-fluctuation-induced nematicity, the accompanying strong symmetry reduction and the significant changes in the lattice parameters together with the opposite sign of the TE anomalies at $T_{\rm S,1}$ and $T_{\rm S,2}$ speak for an additional structural component. Thus, many of the above mentioned experimental findings are consistent with charge/orbital-fluctuation-induced nematicity.\@ Further experiments, however, like doping-dependent diffuse scattering, inelastic x-ray scattering, or high-pressure XRD on Ni122,\@ BaNi$_2$(As,P)$_2$,\@ and Ba(Ni,Co)$_2$As$_2$ are imperative to clarify the nematic character of these systems and the interrelation between the two types of charge density waves.\@

\bibliographystyle{apsrev}
\bibliography{BaNi122_4}

\begin{thebibliography}{30}
\expandafter\ifx\csname natexlab\endcsname\relax\def\natexlab#1{#1}\fi
\expandafter\ifx\csname bibnamefont\endcsname\relax
  \def\bibnamefont#1{#1}\fi
\expandafter\ifx\csname bibfnamefont\endcsname\relax
  \def\bibfnamefont#1{#1}\fi
\expandafter\ifx\csname citenamefont\endcsname\relax
  \def\citenamefont#1{#1}\fi
\expandafter\ifx\csname url\endcsname\relax
  \def\url#1{\texttt{#1}}\fi
\expandafter\ifx\csname urlprefix\endcsname\relax\def\urlprefix{URL }\fi
\providecommand{\bibinfo}[2]{#2}
\providecommand{\eprint}[2][]{\url{#2}}

\bibitem[{\citenamefont{Kamihara et~al.}(2008)\citenamefont{Kamihara, Watanabe,
  Hirano, and Hosono}}]{Kamihara2008}
\bibinfo{author}{\bibfnamefont{Y.}~\bibnamefont{Kamihara}},
  \bibinfo{author}{\bibfnamefont{T.}~\bibnamefont{Watanabe}},
  \bibinfo{author}{\bibfnamefont{M.}~\bibnamefont{Hirano}}, \bibnamefont{and}
  \bibinfo{author}{\bibfnamefont{H.}~\bibnamefont{Hosono}},
  \bibinfo{journal}{Journal of the American Chemical Society}
  \textbf{\bibinfo{volume}{130}}, \bibinfo{pages}{3296} (\bibinfo{year}{2008}).

\bibitem[{\citenamefont{Kim et~al.}(2011)\citenamefont{Kim, Fernandes,
  Kreyssig, Kim, Thaler, Bud'ko, Canfield, McQueeney, Schmalian, and
  Goldman}}]{Kim2011a}
\bibinfo{author}{\bibfnamefont{M.~G.} \bibnamefont{Kim}},
  \bibinfo{author}{\bibfnamefont{R.~M.} \bibnamefont{Fernandes}},
  \bibinfo{author}{\bibfnamefont{A.}~\bibnamefont{Kreyssig}},
  \bibinfo{author}{\bibfnamefont{J.~W.} \bibnamefont{Kim}},
  \bibinfo{author}{\bibfnamefont{A.}~\bibnamefont{Thaler}},
  \bibinfo{author}{\bibfnamefont{S.~L.} \bibnamefont{Bud'ko}},
  \bibinfo{author}{\bibfnamefont{P.~C.} \bibnamefont{Canfield}},
  \bibinfo{author}{\bibfnamefont{R.~J.} \bibnamefont{McQueeney}},
  \bibinfo{author}{\bibfnamefont{J.}~\bibnamefont{Schmalian}},
  \bibnamefont{and} \bibinfo{author}{\bibfnamefont{A.~I.}
  \bibnamefont{Goldman}}, \bibinfo{journal}{Phys. Rev. B}
  \textbf{\bibinfo{volume}{83}}, \bibinfo{pages}{134522}
  (\bibinfo{year}{2011}).

\bibitem[{\citenamefont{Fernandes et~al.}(2014)\citenamefont{Fernandes,
  Chubukov, and Schmalian}}]{Fernandes2014}
\bibinfo{author}{\bibfnamefont{R.~M.} \bibnamefont{Fernandes}},
  \bibinfo{author}{\bibfnamefont{A.~V.} \bibnamefont{Chubukov}},
  \bibnamefont{and}
  \bibinfo{author}{\bibfnamefont{J.}~\bibnamefont{Schmalian}},
  \bibinfo{journal}{Nat. Phys.} \textbf{\bibinfo{volume}{10}},
  \bibinfo{pages}{97} (\bibinfo{year}{2014}).

\bibitem[{foo(2020{\natexlab{a}})}]{footnote19}
\bibinfo{journal}{Sometimes it is discussed that charge and orbital effects
  contribute here to a certain degree as well.}
  (\bibinfo{year}{2020}{\natexlab{a}}).

\bibitem[{\citenamefont{Rotter et~al.}(2008)\citenamefont{Rotter, Tegel, and
  Johrendt}}]{Rotter2008b}
\bibinfo{author}{\bibfnamefont{M.}~\bibnamefont{Rotter}},
  \bibinfo{author}{\bibfnamefont{M.}~\bibnamefont{Tegel}}, \bibnamefont{and}
  \bibinfo{author}{\bibfnamefont{D.}~\bibnamefont{Johrendt}},
  \bibinfo{journal}{Phys. Rev. Lett.} \textbf{\bibinfo{volume}{101}},
  \bibinfo{pages}{107006} (\bibinfo{year}{2008}).

\bibitem[{\citenamefont{Canfield and Bud'ko}(2010)}]{Canfield2010}
\bibinfo{author}{\bibfnamefont{P.~C.} \bibnamefont{Canfield}} \bibnamefont{and}
  \bibinfo{author}{\bibfnamefont{S.~L.} \bibnamefont{Bud'ko}},
  \bibinfo{journal}{Annual Review of Condensed Matter Physics}
  \textbf{\bibinfo{volume}{1}}, \bibinfo{pages}{27} (\bibinfo{year}{2010}).

\bibitem[{\citenamefont{Colombier et~al.}(2009)\citenamefont{Colombier, Bud'ko,
  Ni, and Canfield}}]{Colombier2009}
\bibinfo{author}{\bibfnamefont{E.}~\bibnamefont{Colombier}},
  \bibinfo{author}{\bibfnamefont{S.~L.} \bibnamefont{Bud'ko}},
  \bibinfo{author}{\bibfnamefont{N.}~\bibnamefont{Ni}}, \bibnamefont{and}
  \bibinfo{author}{\bibfnamefont{P.~C.} \bibnamefont{Canfield}},
  \bibinfo{journal}{Phys. Rev. B} \textbf{\bibinfo{volume}{79}},
  \bibinfo{pages}{224518} (\bibinfo{year}{2009}).

\bibitem[{\citenamefont{Sefat et~al.}(2009)\citenamefont{Sefat, McGuire, Jin,
  Sales, Mandrus, Ronning, Bauer, and Mozharivskyj}}]{Sefat2009}
\bibinfo{author}{\bibfnamefont{A.~S.} \bibnamefont{Sefat}},
  \bibinfo{author}{\bibfnamefont{M.~A.} \bibnamefont{McGuire}},
  \bibinfo{author}{\bibfnamefont{R.}~\bibnamefont{Jin}},
  \bibinfo{author}{\bibfnamefont{B.~C.} \bibnamefont{Sales}},
  \bibinfo{author}{\bibfnamefont{D.}~\bibnamefont{Mandrus}},
  \bibinfo{author}{\bibfnamefont{F.}~\bibnamefont{Ronning}},
  \bibinfo{author}{\bibfnamefont{E.~D.} \bibnamefont{Bauer}}, \bibnamefont{and}
  \bibinfo{author}{\bibfnamefont{Y.}~\bibnamefont{Mozharivskyj}},
  \bibinfo{journal}{Phys. Rev. B} \textbf{\bibinfo{volume}{79}},
  \bibinfo{pages}{094508} (\bibinfo{year}{2009}).

\bibitem[{\citenamefont{Kudo et~al.}(2012)\citenamefont{Kudo, Takasuga,
  Okamoto, Hiroi, and Nohara}}]{Kudo2012}
\bibinfo{author}{\bibfnamefont{K.}~\bibnamefont{Kudo}},
  \bibinfo{author}{\bibfnamefont{M.}~\bibnamefont{Takasuga}},
  \bibinfo{author}{\bibfnamefont{Y.}~\bibnamefont{Okamoto}},
  \bibinfo{author}{\bibfnamefont{Z.}~\bibnamefont{Hiroi}}, \bibnamefont{and}
  \bibinfo{author}{\bibfnamefont{M.}~\bibnamefont{Nohara}},
  \bibinfo{journal}{Phys. Rev. Lett.} \textbf{\bibinfo{volume}{109}},
  \bibinfo{pages}{097002} (\bibinfo{year}{2012}).

\bibitem[{\citenamefont{Lee et~al.}(2019)\citenamefont{Lee, de~la Pe\~na, Sun,
  Mitrano, Fang, Jang, Lee, Eckberg, Campbell, Collini et~al.}}]{Lee2019}
\bibinfo{author}{\bibfnamefont{S.}~\bibnamefont{Lee}},
  \bibinfo{author}{\bibfnamefont{G.}~\bibnamefont{de~la Pe\~na}},
  \bibinfo{author}{\bibfnamefont{S.~X.-L.} \bibnamefont{Sun}},
  \bibinfo{author}{\bibfnamefont{M.}~\bibnamefont{Mitrano}},
  \bibinfo{author}{\bibfnamefont{Y.}~\bibnamefont{Fang}},
  \bibinfo{author}{\bibfnamefont{H.}~\bibnamefont{Jang}},
  \bibinfo{author}{\bibfnamefont{J.-S.} \bibnamefont{Lee}},
  \bibinfo{author}{\bibfnamefont{C.}~\bibnamefont{Eckberg}},
  \bibinfo{author}{\bibfnamefont{D.}~\bibnamefont{Campbell}},
  \bibinfo{author}{\bibfnamefont{J.}~\bibnamefont{Collini}},
  \bibnamefont{et~al.}, \bibinfo{journal}{Phys. Rev. Lett.}
  \textbf{\bibinfo{volume}{122}}, \bibinfo{pages}{147601}
  (\bibinfo{year}{2019}).

\bibitem[{\citenamefont{Kudo et~al.}(2017)\citenamefont{Kudo, Takasuga, and
  Nohara}}]{Kudo2017}
\bibinfo{author}{\bibfnamefont{K.}~\bibnamefont{Kudo}},
  \bibinfo{author}{\bibfnamefont{M.}~\bibnamefont{Takasuga}}, \bibnamefont{and}
  \bibinfo{author}{\bibfnamefont{M.}~\bibnamefont{Nohara}},
  \bibinfo{journal}{arXiv:1704.04854}  (\bibinfo{year}{2017}).

\bibitem[{\citenamefont{Eckberg et~al.}(2020)\citenamefont{Eckberg, Campbell,
  Metz, Collini, Hodovanets, Drye, Zavalij, Christensen, Fernandes, Lee
  et~al.}}]{Eckberg2019}
\bibinfo{author}{\bibfnamefont{C.}~\bibnamefont{Eckberg}},
  \bibinfo{author}{\bibfnamefont{D.~J.} \bibnamefont{Campbell}},
  \bibinfo{author}{\bibfnamefont{T.}~\bibnamefont{Metz}},
  \bibinfo{author}{\bibfnamefont{J.}~\bibnamefont{Collini}},
  \bibinfo{author}{\bibfnamefont{H.}~\bibnamefont{Hodovanets}},
  \bibinfo{author}{\bibfnamefont{T.}~\bibnamefont{Drye}},
  \bibinfo{author}{\bibfnamefont{P.}~\bibnamefont{Zavalij}},
  \bibinfo{author}{\bibfnamefont{M.~H.} \bibnamefont{Christensen}},
  \bibinfo{author}{\bibfnamefont{R.~M.} \bibnamefont{Fernandes}},
  \bibinfo{author}{\bibfnamefont{S.}~\bibnamefont{Lee}}, \bibnamefont{et~al.},
  \bibinfo{journal}{Nat.\@ Phys.\@} \textbf{\bibinfo{volume}{16}},
  \bibinfo{pages}{346} (\bibinfo{year}{2020}).

\bibitem[{\citenamefont{Yamakawa et~al.}(2013)\citenamefont{Yamakawa, Onari,
  and Kontani}}]{Yamakawa2013}
\bibinfo{author}{\bibfnamefont{Y.}~\bibnamefont{Yamakawa}},
  \bibinfo{author}{\bibfnamefont{S.}~\bibnamefont{Onari}}, \bibnamefont{and}
  \bibinfo{author}{\bibfnamefont{H.}~\bibnamefont{Kontani}},
  \bibinfo{journal}{Journal of the Physical Society of Japan}
  \textbf{\bibinfo{volume}{82}}, \bibinfo{pages}{094704}
  (\bibinfo{year}{2013}).

\bibitem[{\citenamefont{Merz et~al.}(2016)\citenamefont{Merz, Schweiss, Nagel,
  Huang, Eder, Wolf, von Löhneysen, and Schuppler}}]{Merz2016}
\bibinfo{author}{\bibfnamefont{M.}~\bibnamefont{Merz}},
  \bibinfo{author}{\bibfnamefont{P.}~\bibnamefont{Schweiss}},
  \bibinfo{author}{\bibfnamefont{P.}~\bibnamefont{Nagel}},
  \bibinfo{author}{\bibfnamefont{M.-J.} \bibnamefont{Huang}},
  \bibinfo{author}{\bibfnamefont{R.}~\bibnamefont{Eder}},
  \bibinfo{author}{\bibfnamefont{T.}~\bibnamefont{Wolf}},
  \bibinfo{author}{\bibfnamefont{H.}~\bibnamefont{von Löhneysen}},
  \bibnamefont{and}
  \bibinfo{author}{\bibfnamefont{S.}~\bibnamefont{Schuppler}},
  \bibinfo{journal}{Journal of the Physical Society of Japan}
  \textbf{\bibinfo{volume}{85}}, \bibinfo{pages}{044707}
  (\bibinfo{year}{2016}).

\bibitem[{\citenamefont{Meingast et~al.}(1990)\citenamefont{Meingast, Blank,
  B\"urkle, Obst, Wolf, W\"uhl, Selvamanickam, and Salama}}]{Meingast1990}
\bibinfo{author}{\bibfnamefont{C.}~\bibnamefont{Meingast}},
  \bibinfo{author}{\bibfnamefont{B.}~\bibnamefont{Blank}},
  \bibinfo{author}{\bibfnamefont{H.}~\bibnamefont{B\"urkle}},
  \bibinfo{author}{\bibfnamefont{B.}~\bibnamefont{Obst}},
  \bibinfo{author}{\bibfnamefont{T.}~\bibnamefont{Wolf}},
  \bibinfo{author}{\bibfnamefont{H.}~\bibnamefont{W\"uhl}},
  \bibinfo{author}{\bibfnamefont{V.}~\bibnamefont{Selvamanickam}},
  \bibnamefont{and} \bibinfo{author}{\bibfnamefont{K.}~\bibnamefont{Salama}},
  \bibinfo{journal}{Phys. Rev. B} \textbf{\bibinfo{volume}{41}},
  \bibinfo{pages}{11299} (\bibinfo{year}{1990}).

\bibitem[{\citenamefont{Merz et~al.}(2010)\citenamefont{Merz, Nagel, Pinta,
  Samartsev, v.~L\"ohneysen, Wissinger, Uebe, Assmann, Fuchs, and
  Schuppler}}]{Merz2010}
\bibinfo{author}{\bibfnamefont{M.}~\bibnamefont{Merz}},
  \bibinfo{author}{\bibfnamefont{P.}~\bibnamefont{Nagel}},
  \bibinfo{author}{\bibfnamefont{C.}~\bibnamefont{Pinta}},
  \bibinfo{author}{\bibfnamefont{A.}~\bibnamefont{Samartsev}},
  \bibinfo{author}{\bibfnamefont{H.}~\bibnamefont{v.~L\"ohneysen}},
  \bibinfo{author}{\bibfnamefont{M.}~\bibnamefont{Wissinger}},
  \bibinfo{author}{\bibfnamefont{S.}~\bibnamefont{Uebe}},
  \bibinfo{author}{\bibfnamefont{A.}~\bibnamefont{Assmann}},
  \bibinfo{author}{\bibfnamefont{D.}~\bibnamefont{Fuchs}}, \bibnamefont{and}
  \bibinfo{author}{\bibfnamefont{S.}~\bibnamefont{Schuppler}},
  \bibinfo{journal}{Phys. Rev. B} \textbf{\bibinfo{volume}{82}},
  \bibinfo{pages}{174416} (\bibinfo{year}{2010}).

\bibitem[{\citenamefont{Merz et~al.}(2011)\citenamefont{Merz, Fuchs, Assmann,
  Uebe, v.~L\"ohneysen, Nagel, and Schuppler}}]{Merz2011}
\bibinfo{author}{\bibfnamefont{M.}~\bibnamefont{Merz}},
  \bibinfo{author}{\bibfnamefont{D.}~\bibnamefont{Fuchs}},
  \bibinfo{author}{\bibfnamefont{A.}~\bibnamefont{Assmann}},
  \bibinfo{author}{\bibfnamefont{S.}~\bibnamefont{Uebe}},
  \bibinfo{author}{\bibfnamefont{H.}~\bibnamefont{v.~L\"ohneysen}},
  \bibinfo{author}{\bibfnamefont{P.}~\bibnamefont{Nagel}}, \bibnamefont{and}
  \bibinfo{author}{\bibfnamefont{S.}~\bibnamefont{Schuppler}},
  \bibinfo{journal}{Phys. Rev. B} \textbf{\bibinfo{volume}{84}},
  \bibinfo{pages}{014436} (\bibinfo{year}{2011}).

\bibitem[{\citenamefont{Merz et~al.}(2012)\citenamefont{Merz, Eilers, Wolf,
  Nagel, v.~L\"ohneysen, and Schuppler}}]{Merz2012}
\bibinfo{author}{\bibfnamefont{M.}~\bibnamefont{Merz}},
  \bibinfo{author}{\bibfnamefont{F.}~\bibnamefont{Eilers}},
  \bibinfo{author}{\bibfnamefont{T.}~\bibnamefont{Wolf}},
  \bibinfo{author}{\bibfnamefont{P.}~\bibnamefont{Nagel}},
  \bibinfo{author}{\bibfnamefont{H.}~\bibnamefont{v.~L\"ohneysen}},
  \bibnamefont{and}
  \bibinfo{author}{\bibfnamefont{S.}~\bibnamefont{Schuppler}},
  \bibinfo{journal}{Phys. Rev. B} \textbf{\bibinfo{volume}{86}},
  \bibinfo{pages}{104503} (\bibinfo{year}{2012}).

\bibitem[{\citenamefont{B\"ohmer et~al.}(2015)\citenamefont{B\"ohmer, Hardy,
  Wang, Wolf, Schweiss, and Meingast}}]{Boehmer2015}
\bibinfo{author}{\bibfnamefont{A.~E.} \bibnamefont{B\"ohmer}},
  \bibinfo{author}{\bibfnamefont{F.}~\bibnamefont{Hardy}},
  \bibinfo{author}{\bibfnamefont{L.}~\bibnamefont{Wang}},
  \bibinfo{author}{\bibfnamefont{T.}~\bibnamefont{Wolf}},
  \bibinfo{author}{\bibfnamefont{P.}~\bibnamefont{Schweiss}}, \bibnamefont{and}
  \bibinfo{author}{\bibfnamefont{C.}~\bibnamefont{Meingast}},
  \bibinfo{journal}{Nature Communications} \textbf{\bibinfo{volume}{6}},
  \bibinfo{pages}{7911} (\bibinfo{year}{2015}).

\bibitem[{\citenamefont{Wang et~al.}(2016)\citenamefont{Wang, Hardy, B\"ohmer,
  Wolf, Schweiss, and Meingast}}]{Wang2016}
\bibinfo{author}{\bibfnamefont{L.}~\bibnamefont{Wang}},
  \bibinfo{author}{\bibfnamefont{F.}~\bibnamefont{Hardy}},
  \bibinfo{author}{\bibfnamefont{A.~E.} \bibnamefont{B\"ohmer}},
  \bibinfo{author}{\bibfnamefont{T.}~\bibnamefont{Wolf}},
  \bibinfo{author}{\bibfnamefont{P.}~\bibnamefont{Schweiss}}, \bibnamefont{and}
  \bibinfo{author}{\bibfnamefont{C.}~\bibnamefont{Meingast}},
  \bibinfo{journal}{Phys. Rev. B} \textbf{\bibinfo{volume}{93}},
  \bibinfo{pages}{014514} (\bibinfo{year}{2016}).

\bibitem[{\citenamefont{Wang et~al.}(2018)\citenamefont{Wang, He, Hardy,
  Adelmann, Wolf, Merz, Schweiss, and Meingast}}]{Wang2018}
\bibinfo{author}{\bibfnamefont{L.}~\bibnamefont{Wang}},
  \bibinfo{author}{\bibfnamefont{M.}~\bibnamefont{He}},
  \bibinfo{author}{\bibfnamefont{F.}~\bibnamefont{Hardy}},
  \bibinfo{author}{\bibfnamefont{P.}~\bibnamefont{Adelmann}},
  \bibinfo{author}{\bibfnamefont{T.}~\bibnamefont{Wolf}},
  \bibinfo{author}{\bibfnamefont{M.}~\bibnamefont{Merz}},
  \bibinfo{author}{\bibfnamefont{P.}~\bibnamefont{Schweiss}}, \bibnamefont{and}
  \bibinfo{author}{\bibfnamefont{C.}~\bibnamefont{Meingast}},
  \bibinfo{journal}{Phys. Rev. B} \textbf{\bibinfo{volume}{97}},
  \bibinfo{pages}{224518} (\bibinfo{year}{2018}).

\bibitem[{foo(2020{\natexlab{b}})}]{footnote13}
\bibinfo{journal}{Reducing the symmetry further to monoclinic will not
  significantly change the results derived for the lattice parameters, the
  atomic positions, and the bond distances and angles. While a symmetry
  reduction to monoclinic cannot completely be ruled out, it is generally
  accepted convention to use the higher symmetry in cases such as this, where
  the orthorhombic refinement is merely reproduced by the monoclinic one
  (without improving the reliability factors GOF, $R_1$, and $wR_2$).\@ The
  monoclinic refinement then only comes at the cost of introducing more degrees
  of freedom and, thus, more parameters.}
  (\bibinfo{year}{2020}{\natexlab{b}}).

\bibitem[{foo(2020{\natexlab{c}})}]{footnote14}
\bibinfo{journal}{This finding explains why the tetragonal symmetry is broken
  by magnetic fluctuations in the ${\rm B_{2g}}$ channel for Fe122 while it is
  broken by charge fluctuations in the ${\rm B_{1g}}$ channel for
  (Ba,Sr)Ni$_2$As$_2$ as assumed in \cite{Eckberg2019}.\@}
  (\bibinfo{year}{2020}{\natexlab{c}}).

\bibitem[{\citenamefont{Streltsov and Khomskii}(2014)}]{Streltsov2014}
\bibinfo{author}{\bibfnamefont{S.~V.} \bibnamefont{Streltsov}}
  \bibnamefont{and} \bibinfo{author}{\bibfnamefont{D.~I.}
  \bibnamefont{Khomskii}}, \bibinfo{journal}{Phys. Rev. B}
  \textbf{\bibinfo{volume}{89}}, \bibinfo{pages}{161112}
  (\bibinfo{year}{2014}).

\bibitem[{foo(2020{\natexlab{d}})}]{footnote16}
\bibinfo{journal}{The splitting of the ($h \pm \frac{1}{3}$ $k$ $l \mp
  \frac{1}{3}$)$_{\rm tet}$ reflections is for the ($h$ $k$ ${\bar 1}$)$_{\rm
  tet}$ plane artificially produced by the fact that the RLs of twins with the
  true triclinic symmetry are projected on the RL with the high-$T$ tetragonal
  symmetry. The true position of the reflections, however, can be determined
  from the ($h$ ${\bar 1}$ $l$)$_{\rm tet}$ plane.}
  (\bibinfo{year}{2020}{\natexlab{d}}).

\bibitem[{foo(2020{\natexlab{e}})}]{footnote15}
\bibinfo{journal}{In the present context, the spectrum can directly be
  interpreted as the unoccupied density of states of the Ni $3d$ orbitals with
  the relevant directional characteristics, thus corresponding to the number of
  (hole-type) charge carriers in these orbitals.}
  (\bibinfo{year}{2020}{\natexlab{e}}).

\bibitem[{\citenamefont{Zhang et~al.}(2014)\citenamefont{Zhang, Harriger, Yin,
  Lv, Wang, Tan, Song, Abernathy, Tian, Egami et~al.}}]{Zhang2014}
\bibinfo{author}{\bibfnamefont{C.}~\bibnamefont{Zhang}},
  \bibinfo{author}{\bibfnamefont{L.~W.} \bibnamefont{Harriger}},
  \bibinfo{author}{\bibfnamefont{Z.}~\bibnamefont{Yin}},
  \bibinfo{author}{\bibfnamefont{W.}~\bibnamefont{Lv}},
  \bibinfo{author}{\bibfnamefont{M.}~\bibnamefont{Wang}},
  \bibinfo{author}{\bibfnamefont{G.}~\bibnamefont{Tan}},
  \bibinfo{author}{\bibfnamefont{Y.}~\bibnamefont{Song}},
  \bibinfo{author}{\bibfnamefont{D.~L.} \bibnamefont{Abernathy}},
  \bibinfo{author}{\bibfnamefont{W.}~\bibnamefont{Tian}},
  \bibinfo{author}{\bibfnamefont{T.}~\bibnamefont{Egami}},
  \bibnamefont{et~al.}, \bibinfo{journal}{Phys. Rev. Lett.}
  \textbf{\bibinfo{volume}{112}}, \bibinfo{pages}{217202}
  (\bibinfo{year}{2014}).

\bibitem[{\citenamefont{Sheldrick}(2008)}]{Sheldrick}
\bibinfo{author}{\bibfnamefont{G.~M.} \bibnamefont{Sheldrick}},
  \bibinfo{journal}{Acta Crystallographica Section A}
  \textbf{\bibinfo{volume}{64}}, \bibinfo{pages}{112} (\bibinfo{year}{2008}).

\bibitem[{\citenamefont{Petricek et~al.}(2014)\citenamefont{Petricek, Dusek,
  and Palatinus}}]{Petricek}
\bibinfo{author}{\bibfnamefont{V.}~\bibnamefont{Petricek}},
  \bibinfo{author}{\bibfnamefont{M.}~\bibnamefont{Dusek}}, \bibnamefont{and}
  \bibinfo{author}{\bibfnamefont{L.}~\bibnamefont{Palatinus}},
  \bibinfo{journal}{Zeitschrift für Kristallographie - Crystalline Materials}
  \textbf{\bibinfo{volume}{229}}, \bibinfo{pages}{345} (\bibinfo{year}{2014}).

\bibitem[{\citenamefont{Fernandes and Schmalian}(2012)}]{Fernandes2012}
\bibinfo{author}{\bibfnamefont{R.~M.} \bibnamefont{Fernandes}}
  \bibnamefont{and}
  \bibinfo{author}{\bibfnamefont{J.}~\bibnamefont{Schmalian}},
  \bibinfo{journal}{Superconductor Science and Technology}
  \textbf{\bibinfo{volume}{25}}, \bibinfo{pages}{084005}
  (\bibinfo{year}{2012}).

\end{thebibliography}

\end{document}